\documentclass[twocolumn,resetfootnote]{aastex701}

\newcommand{\phsm}{\mbox{${\mathrm{ph}}\,{\rm s}^{-1}{\mathrm{m}}^{-2}{\mathrm{\mu m}}^{-1}{\mathrm{arcsec}}^{-2}$}}
\newcommand{\magarcsec}{\mbox{mag\,arcsec$^{-2}$}}

\newcommand{\mum}{\mbox{$\mathrm{\mu m}$}}
\newcommand{\mm}{\mbox{$\mathrm{mm}$}}

\newcommand{\K}{\mbox{K}}
\newcommand{\C}{\mbox{$\degr C$}}

\newcommand{\ugqm}{\mbox{$\mathrm{\mu g\,m^{-3}}$}}

\newcommand{\q}{\mbox{$Q_0$}}

\begin{document}

\title{KDP as a thermal blocking filter \\ Deep near IR observations with a warm narrow band filter}

\author[0000-0002-9024-1662]{Joonas K. M. Viuho}
\affiliation{Cosmic Dawn Center (DAWN)}
\affiliation{Niels Bohr Institute, University of Copenhagen, Jagtvej 155A, 2. floor, 2200 Copenhagen N., Denmark}
\affiliation{Nordic Optical Telescope, Rambla José Ana Fernández Pérez 7, ES-38711 Breña Baja, Spain}
\affiliation{Department of Physics and Astronomy, Aarhus University, Munkegade 120, DK-8000 Aarhus C, Denmark}
\email[show]{joonas.viuho@nbi.ku.dk}
\correspondingauthor{Joonas K. M. Viuho}

\author[0000-0001-6316-9880]{Amanda A. Djupvik}
\affiliation{Nordic Optical Telescope, Rambla José Ana Fernández Pérez 7, ES-38711 Breña Baja, Spain}
\affiliation{Department of Physics and Astronomy, Aarhus University, Munkegade 120, DK-8000 Aarhus C, Denmark}
\email{amanda@not.iac.es}

\author[0000-0002-7106-2781]{Anton N. Sørensen}
\affiliation{Cosmic Dawn Center (DAWN)}
\affiliation{Niels Bohr Institute, University of Copenhagen, Jagtvej 155A, 2. floor, 2200 Copenhagen N., Denmark}
\email{norup@nbi.ku.dk}

\author[0009-0009-0999-9651]{Daniel Kumar}
\affiliation{Ferroperm Optics A/S, Stubbeled 7, Trørød, DK-2950 Vedbæk, Denmark}
\email{dk@optics.dk}

\author[0009-0004-8353-7010]{Peder Steiner}
\affiliation{Ferroperm Optics A/S, Stubbeled 7, Trørød, DK-2950 Vedbæk, Denmark}
\email{dk@optics.dk}

\author[0000-0002-8149-8298]{Johan P. U. Fynbo}
\affiliation{Cosmic Dawn Center (DAWN)}
\affiliation{Niels Bohr Institute, University of Copenhagen, Jagtvej 155A, 2. floor, 2200 Copenhagen N., Denmark}
\email{jfynbo@nbi.ku.dk}

\author[0009-0009-0526-050X]{Sergio Armas}
\affiliation{Nordic Optical Telescope, Rambla José Ana Fernández Pérez 7, ES-38711 Breña Baja, Spain}
\affiliation{Department of Physics and Astronomy, Aarhus University, Munkegade 120, DK-8000 Aarhus C, Denmark}
\email{sap@not.iac.es}

\author[0000-0002-8109-033X]{Michael I. Andersen}
\affiliation{Cosmic Dawn Center (DAWN)}
\affiliation{Niels Bohr Institute, University of Copenhagen, Jagtvej 155A, 2. floor, 2200 Copenhagen N., Denmark}
\email{mia@nbi.ku.dk}

\begin{abstract} 
    Ground-based astronomy suffers from strong atmospheric line- and thermal continuum emission, at the near infrared (NIR, 0.7--1.1\,\mum{}), and short-wave infrared (SWIR, 1.1--2.5\,\mum{}) wavelengths.  The thermal continuum emission increases exponentially towards the red sensitivity cutoff of the state-of-the-art 2.5\,\mum{} cutoff SWIR detectors. Given availability of an optical quality shortpass filter material with strong blocking density in the SWIR, lower cost instrumentation, and higher performance filters could be designed. We demonstrate monopotassium dihydrogen phosphate (KDP, chemical formula \mbox{$\mathrm{KH_2PO_4}$}) as a strong candidate for this purpose. KDP is fully transparent at wavelengths from ultraviolet to 1.3\,\mum{}, but becomes highly opaque at wavelengths $>$2\,\mum{}. We demonstrate on-sky use of KDP by improving performance of a cryogenic broadband filter with known off-band thermal leak, and using a non-cryogenic narrow band filter for deep observation. KDP reduces the sky background by 4.5 magnitudes in the leaky Z-band filter we use. Our 4\,nm wide, central wavelength 1.191\,\mum{} narrowband filter in combination with KDP reduces the sky surface brightness by three magnitudes compared to a J broadband, although the effect of KDP is minor due to high blocking density of the broaband filter. We find a sky surface brightness of 18.5\,\magarcsec{} in our bandpass at 1.191\,\mum{}. KDP is an excellent thermal blocker, when its temperature is maintained above its Curie point at 123\,K. Below Curie point, KDP transforms its crystal structure, degrading its otherwise good imaging properties.
\end{abstract}

\keywords{\uat{Near infrared astronomy}{1093} --- \uat{Astronomical instrumentation}{799} --- \uat{Astronomical optics}{88} --- \uat{Optical filters}{2331} --- \uat{Diffuse radiation}{383}}


\section{Introduction}
The near infrared (NIR) to short-wave infrared (SWIR) wavelength range is ideal for studying the early universe, for example, at redshifts above $\sim6$, the Lyman break falls in this region. Designing highly efficient ground-based instrumentation in this wavelength range is a challenge due to atmospheric and instrumental thermal radiation, and lack of an ideal photo-detector. Silicon has a band gap at 1.14\,eV limiting the use of silicon based detectors to wavelengths shorter than 1.1\,\mum{}. While Germanium detectors with a cutoff at 1.8\,\mum{} exist, they typically exhibit high dark current, making them unusable for faint object photometry and spectroscopy. Mercury-Cadmium-Telluride (HgCdTe) SWIR detectors exist with multiple cutoff wavelengths, but the state-of-art version have the cutoff at 2.5\,\mum{}, and are consequently sensitive to ambient thermal emission. The use of these 2.5\,\mum{} cutoff detectors mandates a fully cryogenic instrument design, leading to increased design complexity and cost, making them viable only for the largest observational facilities.

The use of Potassium Dihydrogen Phosphate (KDP, \mbox{$\mathrm{KH_2PO_4}$}) for blocking thermal radiation has been suggested in the literature: to let the 2.5\,\mum{} cutoff detectors be used without a cryogenic optical train \citep{Amado12}. We have not seen a study applying KDP as a thermal radiation blocking filter in the context of astronomical observations. In this work, we will do exactly that and report on our findings. KDP and its deuterated variant Potassium Dideuterium Phosphate (D-KDP, \mbox{$\mathrm{K_2D_2PO_4}$}) are commonly used non-linear optical crystals in laser frequency modulation. Historically, KDP and D-KDP have been studied extensively in the context of high-power lasers for fusion power \citep{Eimerl87,Yoreo02}. In addition to the D-KDP, KDP has a few other isomorphic crystals, which share similar physical properties.

One of the unique physical properties of KDP is its high absorbance in the SWIR wavelength regime, and very high transmission from UV to 1.3\,\mum{} (Fig.\,\ref{fig:kdpodtrans}), making it an excellent thermal radiation blocking filter. This is notable since only a few optical materials exist that absorb in the thermal infrared, while transmitting in VIR--NIR range \citep{Eimerl87,CRClaser}. Available thermal radiation absorbing glasses do not offer nearly as high optical density (OD) in the SWIR and they also absorb at shorter wavelengths, making them less ideal as short-pass filters. KDP's isomorphic crystals (Table\,\ref{table:isomorphs}) show similar properties of being practically fully transparent at wavelengths shorter than their shortpass cutoff wavelength, and very high optical density at wavelengths longer than their cutoff wavelength. Several immediately interesting use cases can be realized, including:

\begin{itemize}
    \item NIR and SWIR instrumentation where a 2.5\,\mum{} cutoff detector is placed in a cryostat, while the entire optical train is maintained in ambient temperature (similar to current instrumentation with silicon photo-detectors).
    \item Enhanced broad-, and especially, narrowband filter designs for the NIR and SWIR wavelengths as the dielectric coating is not required to block wavelengths $>$2\,\mum{}.
    \item Immediate performance improvement on existing systems suffering from thermal light leaks.
    \item Heat blocking from on-the-line-of-sight warm optical components, such as optical fibers, or silicon waveguide optics.
\end{itemize}

\noindent
In this work, we demonstrate the thermal blocking capabilities by improving photometric performance of existing thermally leaking optical system, and we design a custom warm (ambient temperature) narrow-band (NB) filter to operate in conjunction with the KDP window.

\section{Experimental setup}
We used the Nordic Optical Telescope's (NOT) \mbox{NOTCam} instrument \citep{Abbot00} as a test bed. \mbox{NOTCam} is an imager-spectrograph designed to operate in the SWIR wavelength region including the K band. It has a hemispherical entrance baffle, and a cold Lyot stop limiting off-axis thermal background. The instrument temperature is at 73\,\K{}. \mbox{NOTCam} has an aperture wheel, two filter wheels, a stop wheel, and a grism wheel (Fig.\,\ref{fig:notcam}). We use the filter wheels for mounting the broadband filters, and the aperture wheel for mounting the KDP window. Additionally, the warm narrowband filter was installed in front of \mbox{NOTCam}'s cryostat window in the telescope converging beam, i.e. in fully ambient, non-cryogenic conditions.

\subsection{KDP}\label{sec:kdp}
KDP and its deutered variant, D-KDP, are commonly used for frequency modulation in NIR lasers. Possibly, most well known, KDP and D-KDP have been used in the high power lasers of the National Ignition Facility at Lawrence Livermore National Laboratory \citep{Yoreo02}. KDP can be grown to large size, growing it is relatively quick, and it can be polished to optical quality, making it a low cost and easy to work with optical material. KDP crystals can be doped with different impurities, and research on KDP continues today in the field of solid state dye lasers.

Isomorphic crystals of KDP include Ammonium- (ADP), Rubidium- (RDP) and Cesium- (CDP) Dihydrogen Phosphates, and similarly, Potassium (KDA), Arsenic- (ADA), Rubidium (RDA) and Cesium (CDA) arsenates, and their deuterated counterparts (e.g. Ammonium Dideuterium Phosphate, D-ADP). KDP and its isomorphisms are polymorphic crystals which exhibit interesting dielectric and non-linear optical properties. We will focus on two of their physical properties: the spectral transmission, and the Curie temperature $T_{\mathrm{C}}$.

\begin{figure*}[!ht]
    \gridline{\fig{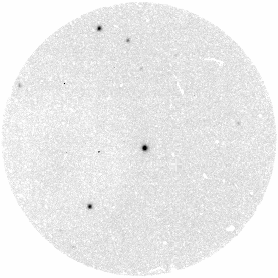}{0.32\textwidth}{(a)}
              \fig{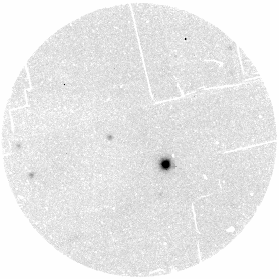}{0.32\textwidth}{(b)}
              \fig{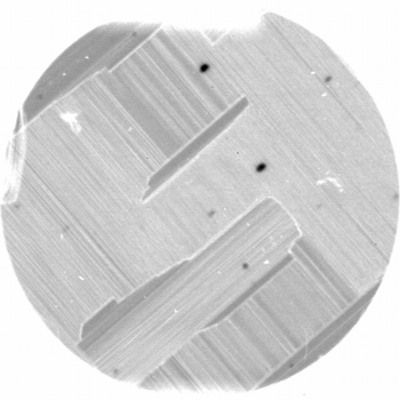}{0.32\textwidth}{(c)}
              }
    \caption{The effect of $T_{\mathrm{C}}$ on the KDP photometric performance. (a) $T_{\mathrm{KDP}}>T_{\mathrm{C}}$ no structure present, (b) $T_{\mathrm{KDP}}\sim T_{\mathrm{C}}$ partial transmission effects, and (c) $T_{\mathrm{KDP}}<T_{\mathrm{C}}$ strong variation in transmission and PSF distortions. The panels (a) and (b) show a 0.5\,mm thick KDP, while the (c) shows a 4\,mm thick KDP.
    \label{fig:kdptc}
    }
\end{figure*}

KDP and its isomorphisms are nearly completely opaque in the thermal infrared, while being fully transparent at wavelengths shorter than 1.3\,\mum{} \citep{Eimerl87}, making them ideal short-pass 'filter glasses', thermal blockers, for NIR filter design. Being polymorphic crystals, KDP and its isomorphisms change from paramagnetic to ferromagnetic state when cooled below their $T_{\mathrm{C}}$. Cooling them below $T_{\mathrm{C}}$ also changes the crystal structure, which in turn introduces unwanted optical qualities: at temperatures below the $T_{\mathrm{C}}$, we see position-dependent transmission non-uniformity and possible polarization-induced point spread function distortions with our z-cut KDPs (Fig.\,\ref{fig:kdptc}). We have observed similar phase-transition effects in both KDP and D-DKP.

Because of the phase transition, KDP and its isomorphs have to be operated above their $T_{\mathrm{C}}$ to maintain the desired optical qualities. In return, this poses a trade-off between the external thermal flux absorbed by the crystal, and the thermal flux emitted by the crystal itself. The $T_{\mathrm{C}}$ is a key property to balance the trade-off. This generally favors materials with lower $T_{\mathrm{C}}$ for SWIR astronomy applications, suggesting the most promising materials to be KDP, KDA and RDA. We summarize the KDP isomorphisms and their Curie temperatures, and approximate short-pass cutoff wavelengths ($\lambda_{\rm SP}$) in Table\,\ref{table:isomorphs}. We refer to the optical and mechanical properties of KDP reported by \cite{Stephenson44a,Zernike64,Eimerl87,CRCchem,CRClaser}.

We purchased three uncoated z-cut KDP and one D-KDP crystals from United Crystals Inc. for this study. Two Ø25\,mm diameter and 4\,mm thick KDPs, and one Ø25\,mm diameter and 4\,mm thick D-KDP were already purchased in 2016. The first observations followed soon after. The phase transition at $T_{\mathrm{C}}$ was not immediately understood, and the project fell dormant for a while for that reason. When it became clear that the change in optical properties was due to the phase transition, calculations were done on the reachable equilibrium temperature inside \mbox{NOTCam} (Sec.\,\ref{sec:eqtemp}). A third Ø10\,mm diameter and 0.5\,mm thick KDP window was purchased and used for the observations reported in this work. 

The KDPs and the D-KDP were installed in \mbox{NOTCam's} aperture wheel (Fig.\,\ref{fig:notcam}). The 4\,\mm{} thick KDPs and the D-KDP were installed in thermal contact with the instrument, while 0.5\,\mm{} thick KDP was prepared with a thermally insulating glass fiber holder. The KDPs in the aperture wheel are placed in the focal plane of the telescope, allowing us to warm the thermally insulated KDP above the $T_{\mathrm{C}}$ by the thermal radiation from the vacuum window (Sec.\,\ref{sec:eqtemp}).

\begin{figure*}[!ht]
    \gridline{\fig{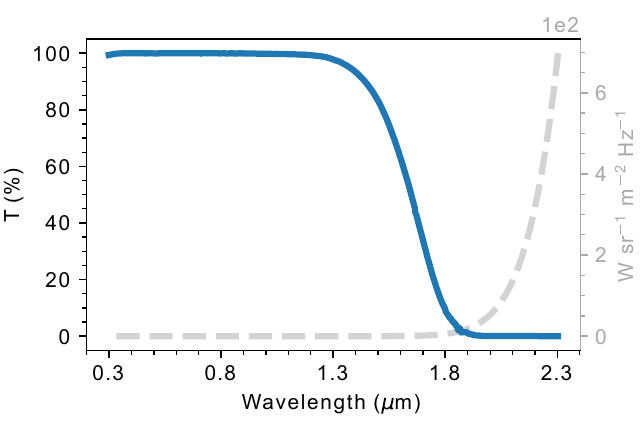}{0.49\textwidth}{(a)}
              \fig{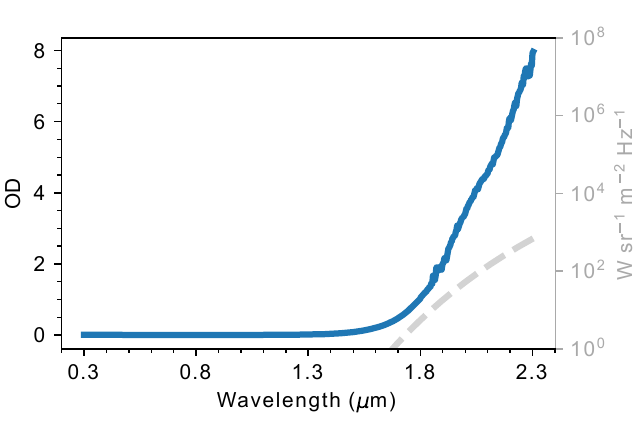}{0.49\textwidth}{(b)}
    }
    \caption{Internal transmission (a) and optical density (b) for unpolarized light through 1\,mm crystal. Gray dashed lines indicate radiance from a $T=288$\,K blackbody.
    \label{fig:kdpodtrans}
    }
\end{figure*}

\begin{deluxetable*}{lccccccccccccc}
    \tabletypesize{\small}
    \caption{Curie temperatures $T_{\mathrm{C}}$ and cut-off wavelengths $\lambda_{\rm SP}$ of KDP and its isomorphisms
    \label{table:isomorphs}
    }   
    \small

    \tablehead{
        \colhead{} & \colhead{KDP} & \colhead{D-KDP} & \colhead{KDA} & \colhead{D-KDA} & \colhead{ADP} & \colhead{D-ADP} & \colhead{ADA} & \colhead{RDP} & \colhead{D-RDP} & \colhead{RDA} & \colhead{D-RDA} & \colhead{CDA} & \colhead{D-CDA}
    }
    \startdata
                                  &       &       &      &       &     &       &     &     &       &     &       &     &       \\
    $T_{\mathrm{C}}$ (K)                     & 123   & 223   & 95   & 159   & 147 & 241   & 216 & 147 & 218   & 110 & 173   & 143 & 241   \\
    $\lambda_{\mathrm{SP,o}}$ (\mum{}) & 1.4   & 1.7   & 1.6\tablenotemark{a} & 1.9\tablenotemark{a} & 1.2\tablenotemark{a} & 1.5   & 1.2 & 1.4 & 1.6   & 1.4 & 1.7   & 1.4 & 1.6   \\
    $\lambda_{\mathrm{SP,e}}$ (\mum{}) & 1.6   & $>$2  &      &       &     & 1.5   & 1.2 & 1.6 & 1.6   & 1.6 & 1.9   & 1.7 & 1.7   \\
                                  &       &       &      &       &     &       &     &     &       &     &       &     &       \\
    \enddata

    \tablenotetext{a}{Unpolarized light.}
    \tablerefs{\citep{Eimerl87}}
    \tablecomments{Curie temperatures $T_{\mathrm{C}}$ and cutoff wavelengths for ordinary $\lambda_{SP,o}$, and extraordinary $\lambda_{SP,e}$ beam have been collected from \cite{Eimerl87}, and are measured for 11\,mm thick samples. We define the $\lambda_{SP}$ as a 50\% falling edge on the transmission curve. D-* indicates the deuterated variants of Potassium (K), Arsenic (A), Rubidium (R), and Cesium (C) phosphates (P), and arsenates (A).}
\end{deluxetable*}

\subsection{The warm narrow-band filter}\label{sec:nbfilter}
Manufacturing high performance interference filters is challenging and expensive, and a trade-off between peak efficiency, in-band transmission, and off-band blocking must be made. In the SWIR, a large spectral range is covered, leaving observations vulnerable to off-band, usually thermal, leaks. The off-band contribution can be significant, especially in narrow-band photometry, where the flux in the band-pass is weak, degrading the photometric performance. Due to the ability of KDP to absorb thermal radiation, filters with band-passes shorter than 1.3\,\mum{} may be placed in ambient (warm) conditions. A KDP thickness of 1\,mm is sufficient to absorb thermal emission from the warm filter, to a level that has no impact on the observations.

\begin{figure*}[!ht]
    \gridline{\fig{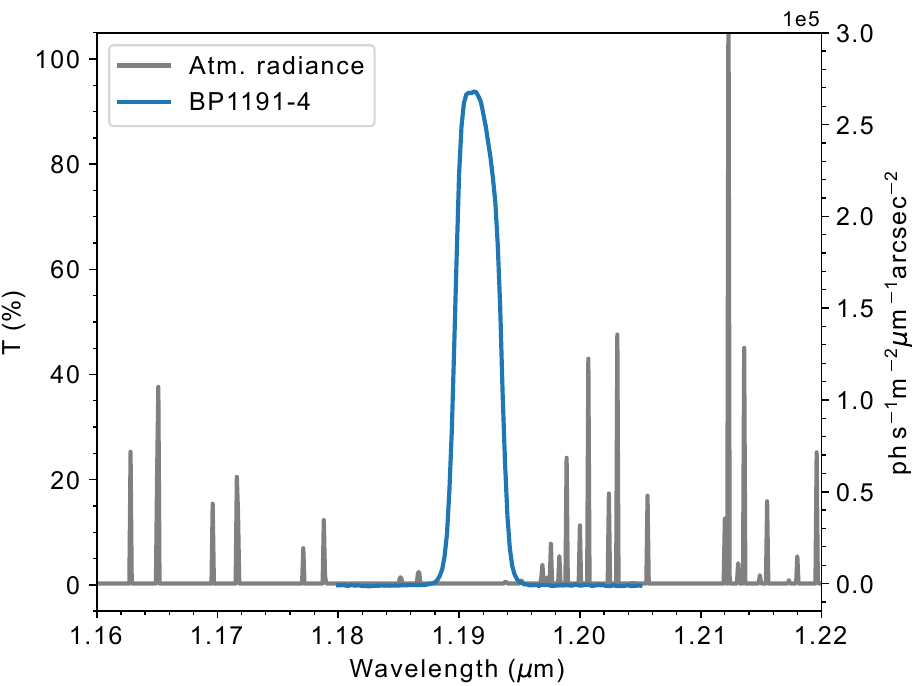}{0.49\textwidth}{(a)}
              \fig{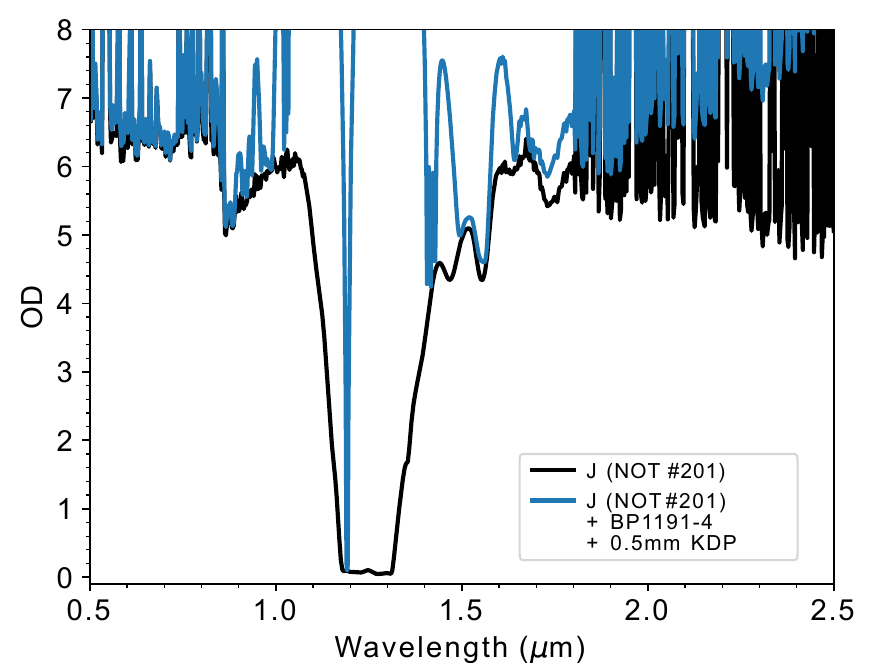}{0.49\textwidth}{(b)}
    }
    \caption{\label{fig:BP1191-4} \mbox{BP1191-4} filter characteristics. Transmission and sky radiance (a), and off-band optical density (b). The bandpass has a 4\,nm full-width-at-half-maximum, and it is centered at 1.191\,\mum{}.}
\end{figure*}

We designed a custom warm NB filter to be used in combination with the \mbox{NOTCam} J-band filter and a 0.5\,mm KDP. The KDP is to ensure that there will be no thermal leak through the broadband filter. The NB filter is only blocking off-band light through the J-band (Fig.\,\ref{fig:BP1191-4}), and it has a central wavelength of 1.191\,\mum{}, with full-width-at-half-maximum of $\sim$4\,nm (R$\sim$300). The bandpass is located in the largest gap between atmospheric hydroxyl (OH) emission lines within the J band. The filter was designed and manufactured by Ferroperm Optics A/S. We will refer it as \mbox{BP1191-4}.

The \mbox{BP1191-4} filter is based on a Fabry-Pérot type design with four cavities having a theoretical bandwidth of 4.2\,nm on the front surface of a Borofloat\,33-substrate. A stress compensating anti-reflection coating was added on the rear surface to increase the transmission. The filter was manufactured in a Generation III Helios 800-chamber, employing the Plasma Assisted Reactive Magnetron Sputtering (PARMS) coating technology. Niobium pentoxide, \mbox{Nb$_2$O$_5$}, was used as a high refractive index and silicon dioxide, \mbox{SiO$_2$}, as a low refractive index material. The monitoring strategy of the bandpass filter was direct monitoring at the center wavelength on a Borofloat\,33-substrate utilizing the self-compensating turning point method \citep{Baumeister04}. The result of the production coating runs was a 1191\,nm filter with a bandwidth of 3.9\,nm and a peak transmission of 94\% measured on a Perkin Elmer L1050 -spectrophotometer (Fig.\,\ref{fig:BP1191-4}). 

\subsection{Broad-band filters}\label{sec:bbfilters}
The \mbox{NOTCam} broadband J, H, K filters were ordered as a part of the Mauna Kea Observatories near-infrared (MKO-NIR) set and have well defined bandpasses \citep{Simons02,Tokunaga02,Tokunaga05}. Two copies of the filters exist at the NOT: one installed inside \mbox{NOTCam}, and another stored in a clean-room outside of the instrument. For this study, we scanned the spare J-band filter, using a Perkin-Elmer L1050 -spectrophotometer at Ferroperm Optics A/S in Vedbæk, Denmark. The spectral range 0.3\,\mum{} to 2.7\,\mum{} was covered, with the motivation of understanding the off-band blocking characteristics of the filter.

\mbox{NOTCam}'s Z- and Y-band filters were intended to have the corresponding UKIDSS bandpasses \citep{Hewett06}. The filters were not manufactured accordingly however, and especially the Z-band filter has poor off-band blocking. The Z and Y band filters were scanned at the Instituto Astrofisica de Canarias (IAC) in range 0.3\,\mum{} to 2.6\,\mum{}. The Z-band filter was confirmed to have a significant thermal leak  (Fig.\,\ref{fig:Zband}), making it an ideal optical component for demonstrating the performance of KDP as a thermal blocker.

\subsection{Detector}\label{sec:detector}
NOTCam is equipped with a 1024 $\times$ 1024 $\times$ 18.5 \mum{} pixel \mbox{HgCdTe/Al$_2$O$_3$} HAWAII PACE I -detector with a cutoff at 2.5\,\mum{} \citep{Kozlowski94}, which is controlled by a Brorfelde Image Acquisition System (BIAS) -controller \citep{Klougart95}. With the NOTCam wide-field camera, the detector covers a \mbox{4\arcmin{} $\times$ 4\arcmin{}} area on sky. The BIAS controller offers two read modes, 1.) reset-read-read (correlated double sampling), and 2.) ramp-sample (non-destructive readout with specified time interval). Due to controller-related specifics not known to us, images from the NOTCam detector show more complex DC structure than other similar detectors. The phenomena is related to the first few reads, and in case of ramp-sampling data can be circumvented by re-fitting the ramps in postprocessing: instead of using the controller provided ramp-sample fits, we have re-fit all data used in this work, ignoring the first read in a sequence.

All four quadrants share the same DC structure, but have slightly differing noise and gain characteristics. Readout noise in a correlated double sampling is 3-4\,ADU, depending on the quadrant. Furthermore, \mbox{NOTCam} shows varying quasiperiodic pickup noise with peak-to-peak amplitude up to $\sim$20\,ADU depending on the exposure. \mbox{NOTCam} dark frames have two distinct components: \emph{i.)} amplifier and shift register glow dependent on number of reads, and \emph{ii.)} an integration time dependent uniform diffuse dark current with rate of \mbox{$\sim$0.01\,ADU\,s$^{-1}$\,px$^{-1}$} over the image area. The register and amplifier glow is shown in Fig.\,\ref{fig:detector}, and it mainly affects areas at the detector edges.

Unlike typical SWIR observations, we are detector noise limited in the shorter integrations in the \mbox{BP1191-4} bandpass. With the \mbox{BP1191-4} filter, we readout the detector in ramp-sampling mode with 14 samples per integration, and in all broadbands, with as many 4\,s sub-reads as it is possible while keeping the total integration within the 1\% linearity limit of the detector. We do not have a measure of the inter-pixel capacitance \citep{Moore06}, and to avoid uncertainty in the gain conversion by determining, and applying, all zeropoints in terms of Analog-to-Digital-Units (ADUs). \mbox{NOTCam} gain measured with the photon transfer tech \citep{Janesick87} is $\sim$2.6\mbox{\,e$^-$\,ADU$^{-1}$}. We use this gain value when estimating the thermal flux from KDP.

\begin{deluxetable*}{cccccl}
    \tablecaption{Observing conditions
    \label{table:weather}
    }                   
    \tablehead{
        \colhead{Night of} & \colhead{Seeing}    & \colhead{Dust}    & \colhead{Temp.}     & \colhead{Rel. humidity} & \colhead{Moon conditions}  \\
        \colhead{ }        & \colhead{\arcsec{}} & \colhead{\ugqm{}} & \colhead{$^\circ$C} & \colhead{\%}            & \colhead{} 
    }
    \startdata
                  &           &         &           &               &        \\
    {2025 Jan 14} & 1.2       & 0.1     & ~9        & 7             & Bright \\
    {2025 Jan 15} & 1.0       & 0.1     & ~5        & $<$1          & Bright \\
    {2025 Jun 30} & 0.5       & 0.1     & 16        & $<$1          & Dark   \\
    {2025 Aug ~8} & 0.6       & 45      & 17        & 38            & Bright \\
    {2025 Aug 23} & 0.9       & 15      & 17        & 75            & Dark   \\
                  &           &         &           &               &        \\
    \enddata
    
\end{deluxetable*}

\section{Observations}\label{sec:observations}
We summarize our observations in Tables \ref{table:weather} and \ref{table:obs}. Initial observations with the 4\,mm thick KDPs were already carried out early in 2016. Results from the observing run were only published on the NOT's website since the data were not clearly understood\footnote{\url{https://www.not.iac.es/instruments/notcam/staff/kdp-tests.html}}. There was a confusion regarding the cause of the changed crystal structure, which had severely affected photometry. At first it was thought that the KDP had fractured during cool down; we only later found out that the structure was due to the phase transition, when the instrument was warm cycled for service. In 2016, the KDP was found to significantly suppress the sky background in the leaky Z-band, prompting further study.

A second observing campaign was started in 2025, after the thermally insulated 0.5\,mm KDP window was installed in \mbox{NOTCam}. Sky conditions on 2025 January 14 and 15 were excellent. New observations were done when the \mbox{BP1191-4} filter became available in 2025 June. The \mbox{BP1191-4} bandpass is blue enough that scattered Moon light would affect observations, thus they have to be obtained under dark sky conditions. The observations on 2025 June 30 were made under excellent conditions. The fields observed in \mbox{BP1191-4} were re-observed in J band on 2025 August 8, but these observations suffered from large amounts of Saharan dust, so called 'Calima'. Since the 2025 June 30 observations were detector systematics limited (Sec.\,\ref{sec:detector}), one more set of \mbox{BP1191-4} observations was made on 2025 August 23, unfortunately also suffering from Saharan dust.

\begin{deluxetable*}{lclcccccccc}
    \tabletypesize{\scriptsize}
    \caption{Table of observations
    \label{table:obs}
    }
    \tablehead{
        \colhead{Standard}     & \colhead{}    & \colhead{Band} & \colhead{Ref.} & \colhead{Date} & \colhead{Start} & \colhead{Integration} & \colhead{Airmass} & \colhead{Aerosols} & \colhead{zp}            & \colhead{Sky surface}  \\
        \colhead{}             & \colhead{}    & \colhead{}     & \colhead{mag.} & \colhead{}     & \colhead{time}  & \colhead{time}        & \colhead{}        & \colhead{total}    & \colhead{ADU\,s$^{-1}$} & \colhead{brightness}   \\
        \colhead{Name}         & \colhead{FS}  & \colhead{}     & \colhead{$m$}  & \colhead{}     & \colhead{UT}    & \colhead{$s$}         & \colhead{$X$}     & \colhead{\ugqm{}}  & \colhead{$m$}           & \colhead{\magarcsec{}}                      
    }
    \startdata
                     &     &                &        &                &        &                        &         &         &                   &              \\
        RL\,149      & 14  & K              & 11.459 & 2025 Jan 15    & 03:00  & 5$\times$(8$\times$4s) & 1.33    & 0.12    & 22.34$\pm$0.00~~~ & 12.48~~      \\
                     &     & H              & 11.438 & 2025 Jan 15    & 03:06  & 5$\times$(8$\times$4s) & 1.35    & 0.12    & 22.92$\pm$0.00~~~ & 13.57~~      \\
                     &     & J              & 11.444 & 2025 Jan 15    & 02:44  & 5$\times$(8$\times$4s) & 1.28    & 0.12    & 22.97$\pm$0.00~~~ & 15.59~~      \\
                     &     & Y              & 11.402 & 2025 Jan 15    & 02:49  & 5$\times$(8$\times$4s) & 1.30    & 0.12    & 22.77$\pm$0.00~~~ & 16.63~~      \\
                     &     & Z              & 11.317 & 2025 Jan 15    & 02:54  & 5$\times$(8$\times$4s) & 1.31    & 0.12    & 22.37$\pm$0.00~~~ & 12.91~~      \\
                     &     &                &        &                &        &                        &         &         &                   &              \\
                     &     & K+KDP          & 11.459 & 2025 Jan 15    & 03:17  & 5$\times$(8$\times$4s) & 1.33    & 0.05    & 15.10$\pm$0.17~~~ & 12.32~~      \\
                     &     & H+KDP          & 11.438 & 2025 Jan 15    & 03:12  & 5$\times$(8$\times$4s) & 1.37    & 0.05    & 22.39$\pm$0.00~~~ & 13.63~~      \\
                     &     & J+KDP          & 11.444 & 2025 Jan 15    & 02:16  & 5$\times$(8$\times$4s) & 1.22    & 0.06    & 22.83$\pm$0.00~~~ & 16.77~~      \\
                     &     & Y+KDP          & 11.402 & 2025 Jan 15    & 02:10  & 5$\times$(8$\times$4s) & 1.21    & 0.06    & 22.51$\pm$0.00~~~ & 16.82~~      \\
                     &     & Z+KDP          & 11.317 & 2025 Jan 15    & 02:04  & 5$\times$(8$\times$4s) & 1.20    & 0.06    & 22.25$\pm$0.00~~~ & 17.42~~      \\
                     &     &                &        &                &        &                        &         &         &                   &              \\
        GD\,71       & 12  & H              & 13.805 & 2025 Jan 16    & 02:06  & 5$\times$(5$\times$4s) & 1.30    & 0.08    & 22.87$\pm$0.01~~~ & 13.87~~      \\
                     &     & J              & 13.710 & 2025 Jan 16    & 02:02  & 5$\times$(5$\times$4s) & 1.28    & 0.08    & 22.90$\pm$0.01~~~ & 15.66~~      \\
                     &     & Y              & 13.657 & 2025 Jan 16    & 01:58  & 5$\times$(5$\times$4s) & 1.27    & 0.08    & 22.65$\pm$0.01~~~ & 17.04~~      \\
                     &     & Z              & 13.398 & 2025 Jan 16    & 01:53  & 5$\times$(5$\times$4s) & 1.25    & 0.08    & 22.32$\pm$0.01~~~ & 13.08~~      \\
                     &     &                &        &                &        &                        &         &         &                   &              \\
                     &     & H+KDP          & 13.805 & 2025 Jan 16    & 01:35  & 5$\times$(5$\times$4s) & 1.19    & 0.08    & 22.42$\pm$0.01~~~ & 14.22~~      \\
                     &     & J+KDP          & 13.710 & 2025 Jan 16    & 01:40  & 5$\times$(5$\times$4s) & 1.21    & 0.08    & 22.81$\pm$0.01~~~ & 15.88~~      \\
                     &     & Y+KDP          & 13.657 & 2025 Jan 16    & 01:45  & 5$\times$(5$\times$4s) & 1.22    & 0.08    & 22.62$\pm$0.01~~~ & 17.43~~      \\
                     &     & Z+KDP          & 13.398 & 2025 Jan 16    & 01:49  & 5$\times$(5$\times$4s) & 1.24    & 0.08    & 22.27$\pm$0.01~~~ & 18.19~~      \\
                     &     &                &        &                &        &                        &         &         &                   &              \\
        GSPC         & --  & H              & 11.585 & 2025 Jan 16    & 00:22  & 5$\times$(4$\times$4s) & 1.13    & 0.05    & 22.92$\pm$0.00~~~ & 14.02~~      \\
        P545-C       &     & J              & 11.841 & 2025 Jan 16    & 00:26  & 5$\times$(4$\times$4s) & 1.14    & 0.05    & 22.97$\pm$0.00~~~ & 15.68~~      \\
                     &     & Y              & 12.031 & 2025 Jan 16    & 00:30  & 5$\times$(4$\times$4s) & 1.15    & 0.05    & 22.63$\pm$0.01~~~ & 16.68~~      \\
                     &     & Z              & 12.049 & 2025 Jan 16    & 00:34  & 5$\times$(4$\times$4s) & 1.15    & 0.05    & 22.26$\pm$0.01~~~ & 13.80~~      \\
                     &     &                &        &                &        &                        &         &         &                   &              \\
                     &     & H+KDP          & 11.585 & 2025 Jan 16    & 00:50  & 5$\times$(4$\times$4s) & 1.18    & 0.05    & 22.51$\pm$0.00~~~ & 14.21~~      \\
                     &     & J+KDP          & 11.841 & 2025 Jan 16    & 00:46  & 5$\times$(4$\times$4s) & 1.18    & 0.05    & 22.85$\pm$0.00~~~ & 15.82~~      \\
                     &     & Y+KDP          & 12.031 & 2025 Jan 16    & 00:42  & 5$\times$(4$\times$4s) & 1.17    & 0.05    & 22.52$\pm$0.01~~~ & 16.91~~      \\
                     &     & Z+KDP          & 12.049 & 2025 Jan 16    & 00:38  & 5$\times$(4$\times$4s) & 1.16    & 0.05    & 22.26$\pm$0.01~~~ & 17.52~~      \\
                     &     &                &        &                &        &                        &         &         &                   &              \\
        GSPC         & 121 & H              & 11.436 & 2025 Jan 16    & 00:54  & 5$\times$(5$\times$4s) & 1.21    & 0.05    & 22.97$\pm$0.00~~~ & 14.08~~      \\
        S772-G       &     & J              & 11.984 & 2025 Jan 16    & 00:59  & 5$\times$(5$\times$4s) & 1.21    & 0.05    & 22.96$\pm$0.00~~~ & 15.81~~      \\
                     &     & Y              & 12.464 & 2025 Jan 16    & 01:20  & 5$\times$(5$\times$4s) & 1.21    & 0.05    & 22.71$\pm$0.01~~~ & 17.21~~      \\
                     &     & Z              & 12.735 & 2025 Jan 16    & 01:06  & 5$\times$(5$\times$4s) & 1.22    & 0.05    & 22.44$\pm$0.01~~~ & 13.24~~      \\
                     &     &                &        &                &        &                        &         &         &                   &              \\
                     &     & H+KDP          & 11.436 & 2025 Jan 16    & 01:24  & 5$\times$(5$\times$4s) & 1.21    & 0.05    & 22.45$\pm$0.00~~~ & 14.21~~      \\
                     &     & J+KDP          & 11.984 & 2025 Jan 16    & 01:20  & 5$\times$(5$\times$4s) & 1.21    & 0.05    & 22.85$\pm$0.00~~~ & 15.89~~      \\
                     &     & Y+KDP          & 12.464 & 2025 Jan 16    & 01:16  & 5$\times$(5$\times$4s) & 1.21    & 0.05    & 22.61$\pm$0.01~~~ & 17.41~~      \\
                     &     & Z+KDP          & 12.735 & 2025 Jan 16    & 01:11  & 5$\times$(5$\times$4s) & 1.22    & 0.05    & 22.28$\pm$0.01~~~ & 18.25~~      \\
                     &     &                &        &                &        &                        &         &         &                   &              \\
        GSPC         & 151 & J              & 12.211 & 2025 Aug ~9    & 00:31  & 5$\times$(4$\times$4s) & 1.01    & 40      & 22.64$\pm$0.01\tablenotemark{a} & 15.66\tablenotemark{b} \\
        P340-H       &     & J+BP1191-4     & 12.211 & 2025 Jul ~~\,1 & 03:01  & 9$\times$(14$\times$5s)& 1.01    & 0.1     & 18.65$\pm$0.04~~~ & 18.52~~      \\
                     &     & J+BP1191-4+KDP & 12.211 & 2025 Jul ~~\,1 & 02:48  & 9$\times$(14$\times$5s)& 1.02    & 0.1     & 18.57$\pm$0.05~~~ & 18.73~~      \\
                     &     & J+BP1191-4     & 12.211 & 2025 Aug 23    & 22:14  & 4$\times$(14$\times$50s)& 1.11    & 21      & 18.79$\pm$0.04\tablenotemark{a} & 18.18\tablenotemark{b} \\
                     &     & J+BP1191-4+KDP & 12.211 & 2025 Aug 23    & 21:24  & 4$\times$(14$\times$50s)& 1.24    & 18      & 18.45$\pm$0.05\tablenotemark{a} & 18.10\tablenotemark{b} \\
                     &     &                &        &                &        &                        &         &         &                   &              \\
        GSPC         & 141 & J              & 11.152 & 2025 Aug ~8    & 22:19  & 5$\times$(4$\times$4s) & 1.01    & 23      & 22.58$\pm$0.00\tablenotemark{a} & 15.69\tablenotemark{b} \\
        P389-D       &     & J+BP1191-4     & 11.152 & 2025 Jul ~~\,1 & 02:25  & 9$\times$(14$\times$5s)& 1.13    & 0.08    & 18.76$\pm$0.02~~~ & 18.54~~      \\
                     &     & J+BP1191-4+KDP & 11.152 & 2025 Jul ~~\,1 & 02:08  & 9$\times$(14$\times$5s)& 1.09    & 0.08    & 18.67$\pm$0.02~~~ & 18.63~~      \\
                     &     &                &        &                &        &                        &         &         &                   &              \\
    \enddata
    \tablenotetext{a}{Dusty observing conditions.} \tablenotetext{b}{Dusty observing conditions, zeropoint applied from another night.}
    \tablerefs{\href{https://gemini.edu/observing/resources/near-ir-resources/photometry/ukirt-standards}{Gemini UKIRT MKO photometric standard stars catalogue}, \citet{Legget06}.}
    \tablecomments{Magnitudes given in the Vega system. The second column indicated the Faint standard (FS) -catalogue number.}
\end{deluxetable*}

\section{Analysis and results}\label{sec:analysis}
\subsection{KDP transmission and optical density}\label{sec:kdpod}
When considering filter transmission curves, transmission is the relevant term in-band, while optical density (OD) is the relevant term off-band. Thus we show graphs of both transmission and optical density. Optical density is related to transmission $T$ via

\begin{equation}
    \mathrm{OD} = -\log_{10}(T)
\end{equation}

\noindent
The optical density of KDP was measured at Ferroperm Optics with the PerkinElmer-L1050 spectrophotometer, at ambient temperature within the wavelength range of 0.300\,\mum{} to 2.315\,\mum{} using unpolarized light. Data from scans of a 4\,mm ($<$1.86\,\mum{}) and a 0.5\,mm ($>$1.86\,\mum{}) thick samples were combined to yield the equivalent optical density of 1\,mm KDP and is presented in Fig.\,\ref{fig:kdpodtrans} and Tables\,\ref{table:od1}, and \ref{table:od2}. The KDP samples were removed from their vacuum bags just before the scan in order not to expose them to humidity. The instrument was null calibrated prior to the measurements. Additionally, the KDP internal transmission was calculated from relative photometry, and is in agreement with the spectrophotometer scans (Table \ref{table:trans}).

\subsection{KDP equilibrium temperature and operating point}\label{sec:eqtemp}
For KDP, $T_{\mathrm{C}}$=123\,\K{}. Consequently, when placed inside \mbox{NOTCam} at $T_{\mathrm{inst}}$=73\,\K{}, KDP undergoes the Curie transition from paramagnetic to ferromagnetic, which also changes its crystal structure. At least, in the case of our z-cut KDPs, the imaging  characteristics of the KDP become worse in the ferromagnetic state. Strong position dependent variability in transmission and point spread function distortion (possibly due to polarization effects) appear below $T_{\mathrm{C}}$, making KDP unusable for high quality photometry, unless it is heated to a temperature above $T_{\mathrm{C}}$.

In order to warm up KDP above its $T_{\mathrm{C}}$, we move it into the telescope beam where it is exposed to ambient thermal radiance. We find the equilibrium temperature of our thermally insulated 0.5\,mm KDP using Stefan-Boltzmann's law, assuming that all energy transfer is radiative. The radiance $L$ of a blackbody given by Stefan-Boltzmann's law is 

\begin{equation}
    L = A\epsilon\sigma T^4 \int\Omega
\end{equation}

\noindent
where $A$ is the surface area, $\epsilon$ the emissivity, $\sigma$ the Stefan-Boltzmann constant, $T$ the temperature of the blackbody, and $\Omega$ the solid angle. On the front side of the KDP, we take typical summer night time ambient temperature $T_{\mathrm{amb}}$=288\,K which arrives from a solid angle visible through NOTCam's entrance baffle, projected through the collimator optics $\Omega_{f,1}$ defined by a cone with semi-angle $\theta$ as 

\begin{equation}
    \Omega_{f,1} = 2\pi\left(1-\cos(\theta) \right)
\end{equation}

\noindent
where the semi-angle $\theta$ is derived from the baffle opening angle, with a diameter of Ø53\,mm and a distance of 68\,mm to the KDP. The remaining half hemisphere $\Omega_{f,2}=2\pi-\Omega_{f,1}$ sees the cold instrument with the cryostat temperature $T_{\mathrm{inst}}$=73\,K. On the rear side, the KDP is exposed to a half hemisphere $\Omega_{r}$ with temperature $T_{\mathrm{inst}}$ (Fig.\,\ref{fig:notcam}). We get the radiant power $P=L/A$, or

\begin{equation}
    \label{eq:power}
    P = \epsilon\sigma\left( T_{\mathrm{amb}}^4\int_{\Omega_{f,1}} d\Omega + T_{\mathrm{inst}}^4\int_{\Omega_{f,2}}d\Omega + T_{\mathrm{inst}}^4\int_{\Omega_{r}}d\Omega \right),
\end{equation}

\noindent
and solve for the equilibrium temperature KDP $T_{\mathrm{KDP}}$ getting,

\begin{equation}
    T_{\mathrm{KDP}} = T_{\mathrm{inst}} + \frac{mP}{c_m(T) \rho VM}\Delta t,
\end{equation}

\noindent 
where $P$ is the total radiant power (Eq.\,\ref{eq:power}), $M$=136.086 molecular weight of KDP, $\rho$=2.338\,\mbox{g\,cm$^-3$} the density of KDP \citep{CRCchem}, and $c_m$ the molar heat capacity as tabulated by \citet{Stephenson44a}. Assuming the emissivity to be unity, we find that it takes about 20--35\,\mbox{min} to warm the 0.5\,\mm{} thermally insulated KDP above its $T_{\mathrm{C}}$, and make the unwanted structure to disappear (Fig.\,\ref{fig:kdpinout}). Given $T_{\mathrm{amb}}$=288\,K we find the KDP to reach an equilibrium temperature of $\sim$150\,\K{} when left in the beam.

\subsection{Data reduction}
Observations consisted of either 5-point or 9-point skewed dither-patterns with 10--15\arcsec{} offsets between two consecutive integrations. All data were recorded as ramp-sampling integrations (Sec.\,\ref{sec:detector}). The broadband standard star fields were recorded with a sample time of 4\,s with a number of samples yielding a total integrated flux within the 1\% linearity range of the detector. The NB integrations had very low flux levels and were readout with the maximum 14 ramp-samples available. All sub-reads are stored, and we re-fit to the sub-reads ignoring the first read in a ramp-sample sequence. In the \mbox{NOTCam} data, the first line and row of each detector quadrant reads out as value 0. These together with hot and cold pixels were removed by bilinear interpolation based on the surrounding pixel values. 

The images were then sky subtracted, aligned, and co-added with {'notcam.cl'} reduction package (priv. comm. A. A. Djupvik, T. Reynolds)\footnote{\url{www.not.iac.es/instruments/notcam}} written for Image Reduction and Analysis Facility (IRAF) \citep{IRAF1,IRAF2}. The first image recorded for each dither-pattern was rejected from the co-addition. For the sky level being very low in the narrowband images, the NB images were additionally co-added without sky subtraction for estimating the sky level, and the instrumental background in case of KDP observations. Isophotal fluxes of the standards stars were extracted, and extinction corrected using the standard extinction coefficients adopted in the NOT's signal-to-noise calculator extinction co-efficients ($\kappa_Z=0.05$, $\kappa_Y=0.07$, $\kappa_J=0.09$, $\kappa_H=0.07$, $\kappa_K=0.08$\,\magarcsec{}). We derive the zeropoints in units of ADU\,s$^{-1}$ from the photometric nights (January 14, January 15, and June 30), and apply them to the sky surface brightness measurement of all nights. We have not derived color terms for the \mbox{BP1191-4} filter, and use the J-band magnitudes for relative photometry in the \mbox{BP1191-4} bandpass.

\subsection{Instrumental background}\label{sec:instrumentalforeground}
The Ø10mm diameter KDP vignettes the \mbox{NOTCam} wide-field camera's 4\arcmin{}$\times$4\arcmin{} field-of-view to a Ø1.1\arcmin{} image circle allowing us to sample detector area that is not exposed to the sky (Fig.\,\ref{fig:detector}). We do not see a difference between the non-illuminated areas in the on-sky KDP exposures shutter open, and the dark exposures shutter closed in the Y- and J-band images. However, with the filters which pass thermal IR Z (thermal leak), H (transmits partially), and K (transmits fully), a light leak from the adjacent open aperture wheel position can be seen in the KDP exposures (Fig.\,\ref{fig:detector}).  We conclude that the black surface treatment is absorbing at the bluer wavelengths, but becomes reflective in the thermal infrared. The fiberglass holder is slightly transparent in the K band (Fig.\,\ref{fig:detector}). The transmission can be seen in the Z- and H-band observations if inspected carefully. Apart from the leak in the aperture wheel, the fiberglass holder, and detector dark current, we do not observe any other instrumental contribution in our KDP integrations, and do not expect any thermal emission from within the instrument to reach the detector.

\subsection{Thermal emission from the KDP}
The KDP itself is a thermal emission source. It is placed in the aperture wheel which is re-imaged on the detector plane. We estimate the KDP to reach a temperature of $\sim$150\,K when placed in the beam. According to our calculations this results to a flux of $\sim$2$\times10^{-4}$\,ADU\,s$^{-1}$\,px$^{-1}$ on the detector when viewed through the F/5.8 wide-field camera. The resulting flux is negligible compared to the sky flux we measure. In order for the KDP to emit a comparable amount to the sky flux, the KDP window would need to reach a temperature of 177\,\K{}, or higher, which we do not consider possible.

\subsection{The effect of persistence and reciprocity failure}\label{sec:persistence}
We do not have a sophisticated model available for correcting persistence and reciprocity failure \citep{Tulloch19}. We do see a power law decay of the background level, in the non-illuminated detector areas in our KDP integrations. We attribute the exponential decay to persistence, and subtract it out as a parasitic signal. Additionally, in case of the long \mbox{BP1191-4} integrations we see spatial correlation in our sky images with the location of the brightest stars. The full field-of-view in the observations without KDP allows us to find a sky area which has not been visited by a star in the last few frames. However, in the integrations with the KDP, the usable image area is significantly smaller (Fig.\,\ref{fig:detector}), restricting the sky are to be measured in the proximity of the field stars. If the effect in the \mbox{BP1191-4} observations with and without KDP is similar, this could lead us to over estimate the sky surface brightness by $\sim$15\% in the narrowband KDP observations.

\begin{figure*}[!ht]
    \gridline{\fig{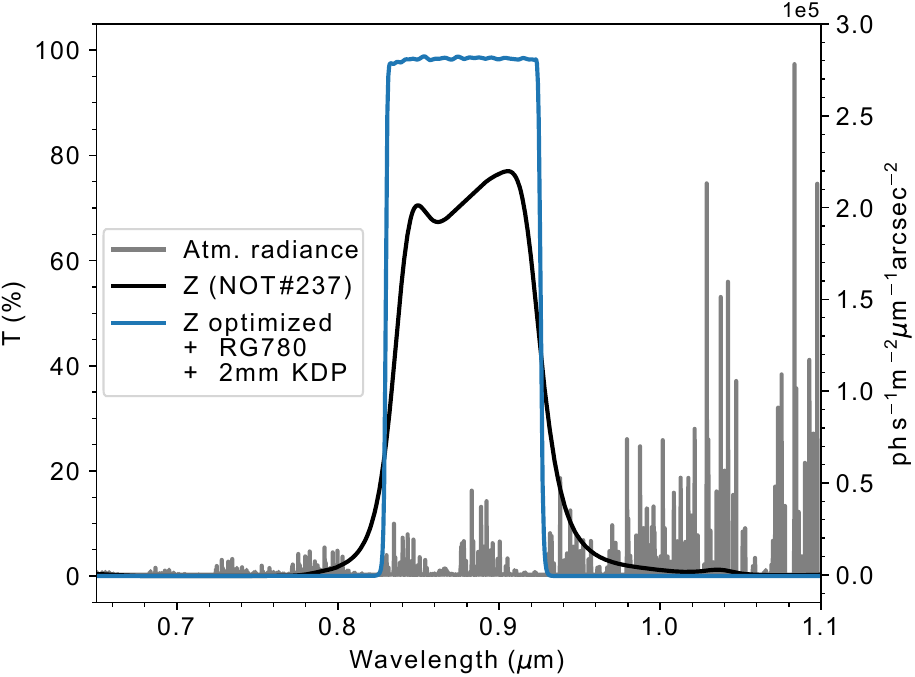}{0.49\textwidth}{(a)}
              \fig{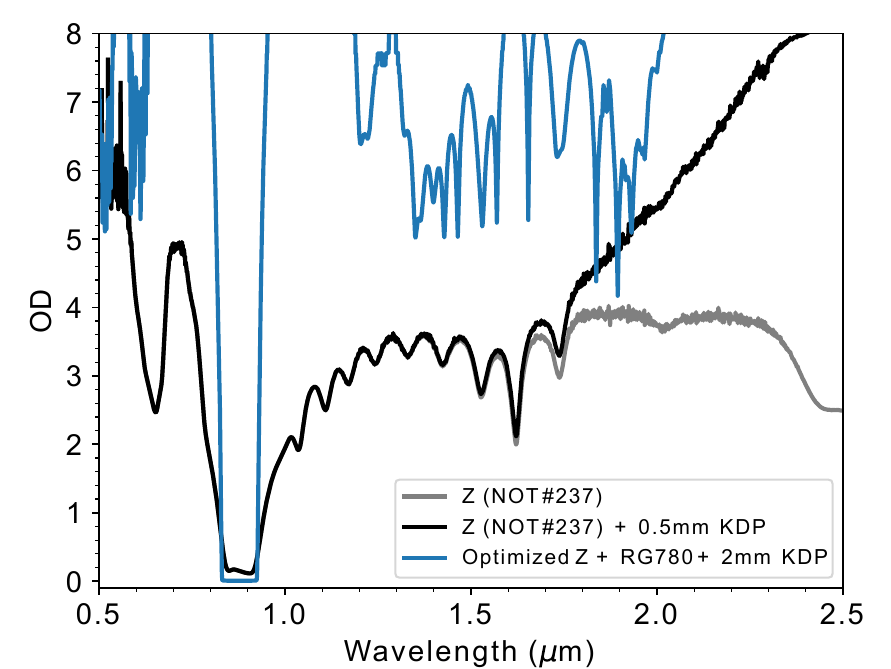}{0.49\textwidth}{(b)}
    }
    \caption{Transmission and off-band blocking of the \mbox{NOTCam} Z band and the \mbox{NOTCam} Z band in combination with KDP used in this study (a). An optimized Z band coating design utilizing KDP as a short pass filter 'glass' is illustrated (b) Optical density of the NOT \#237 Z band without KDP (grey), with KDP (black), and an optimized Z band coating (blue).
    \label{fig:Zband}
    }
\end{figure*}

\subsection{Airglow surface brightness at 1.191\,\mum{}}\label{sec:airglow1191}
Depending on the night, we measure a sky surface brightness of 18.1--18.7\,\magarcsec{}, or 640--1100\,\phsm{} in the \mbox{BP1191-4}. All of the \mbox{BP1191-4} observations have been made under dark Moon conditions. On 2025 June 30, the Moon had set 1.5\,h before the beginning of observations, and on 2025 August 23 was new Moon.

To isolate the atmospheric continuum emission background, we calculate the zodiacal light brightness for our pointings with the InfraRed Science Archive (IRSA), Infrared Processing \& Analysis Center (IPAC) Euclid background model calculator \mbox{Version 1}\footnote{\url{https://irsa.ipac.caltech.edu/applications/BackgroundModel/}}. The calculator is based on the zodiacal light model of \cite{Kelsall98}. Both of the standard star fields observed in the \mbox{BP1191-4} bandpass are located approximately at ecliptic latitude of $b_{\rm ec}$=45$\degr$, and the zodiacal light contributes only 35--40\,\phsm{}, or $\sim$21.8\,\magarcsec{}, which is much less than the sky surface brightness we measure.

In order to estimate the off-band sky light not rejected by our filter setup, and the potential contribution from emission lines in the wings of the \mbox{BP1191-4} bandpass, we take the sky radiance model of ESO SkyCalc \citep{Noll12,Jones13} for La Silla. We do not know the exact quantum efficiency figure for our detector but assume it to be 65\% in the range of 0.5--2.6\,\mum{}. Assuming 95\% reflectivity on the telescope mirrors, 99\% transmission on all lenses, and 7.2\% Fresnel losses on the KDP window, we estimate the sky lines in the wings of the \mbox{BP1191-4} filter bandpass to contribute $\sim$9$\times10^{-4}$ADU\,s$^{-1}$\,px$^{-1}$ to the measured sky surface brightness. Likewise, the integrated sky emission outside of the \mbox{BP1191-4} bandpass would contribute $\sim$1$\times10^{-3}$ADU\,s$^{-1}$\,px$^{-1}$. The transmission and reflectivity figures are optimistic and should give an upper limit on the excess sky contribution. We conclude that we do not see significant amount of unwanted light in our measurement.

\cite{Nguyen16} has measured 18.8\,\magarcsec{} in a bandpass similar to ours at the Table Mountain Observatory, Big Pines, California. Other narrowband measurements in a similar bandpass, have measured a brighter sky surface brightness of $\sim$17.5\,\magarcsec{} \citep[see][for a compilation]{Milvang13}. Spectroscopic observations of \cite{Sullivan12} at Las Campanas, Chile indicate similar values, with a median dark time sky continuum brightness of 18.6\,\magarcsec{}. \cite{Noll24} gives a yearly average continuum background of 230\,\phsm{}, or 19.8\,\magarcsec{} at 1.191\,\mum{} for Cerro Paranal, Chile. 

\begin{deluxetable*}{lcccccc}
    \tabletypesize{\small}
    \caption{Average sky surface brightness in our observations, and the KDP window transmission in the bandpasses
    \label{table:trans}
    }
    \tablehead{
        \colhead{} & \colhead{Z}     & \colhead{Y}     & \colhead{J}     & \colhead{H}     & \colhead{K}     & \colhead{J+BP1191-4} \\
        \colhead{} & \colhead{\#237} & \colhead{\#236} & \colhead{\#201} & \colhead{\#203} & \colhead{\#207} & \colhead{}                 
    }
    \startdata
                    &              &              &              &              &              &                  \\
    Sky with KDP    & 17.8         & 17.1         & 16.1         & 14.1         & 12.3         & 18.5             \\
    Sky without KDP & 13.3         & 16.9         & 15.6         & 13.9         & 12.5         & 18.4             \\
                    &              &              &              &              &              &                  \\
    Transmission    & 92\%         & 89\%         & 91\%         & 64\%         & $<$0.2\%     & 92\%             \\
                    &              &              &              &              &              &                  \\
    \enddata

\tablecomments{Surface brightnesses given in \magarcsec{} (Vega). The KDP windows have not been anti-reflection coated, and the reported transmission includes reflection losses totaling 7.4\,\% on the two surfaces. The large difference in the Z band sky brightness is expected due to the large amount of red leak in the NOT \#237 filter which is blocked by the KDP. 
}
\end{deluxetable*}


\section{Discussion}\label{sec:discussion}

\subsection{KDP as a thermal blocking filter}\label{sec:KDPblocking}
We have demonstrated that KDP can be deployed effectively to reduce unwanted thermal leaks. We achieve a reduction of 4.5\,magnitudes in the \mbox{NOTCam} Z band sky surface brightness comparing observation with and without the 0.5\,mm thick KDP  (Table\,\ref{table:obs}, Fig.\,\ref{fig:zcomp}). The reduction is significant and illustrates the potential of KDP for thermal blocking, or heat absorbing, purposes in astronomical instrumentation. We do not see thermal emission contribution from our warm NB filter. This opens a possibility to place NB filters outside of the cryostat, allowing exchanging them without breaking the cryo-vacuum.

\subsection{Possibilities in filter design}
KDP and its isomorphisms open up new possibilities for NIR filter design since they can be used as a shortpass filter 'glass'. We have designed and demonstrated use of a highly efficient narrowband filter operating outside the instrument cryostat (Fig.\,\ref{fig:BP1191-4}). Furthermore, we show a design for a highly optimized 'blue' NIR filter to be used with a 2.5\,\mum{} cutoff HgCdTe detector. The implementation of the Z band (UKIDSS Z\footnote{\url{http://www.ukidss.org/technical/instrument/filters.html}}, \citet{Hewett06}) allows high peak transmission, and well defined bandpass edges due to the relaxed blocking requirements of the dielectric coating at wavelengths $>$1.9\,\mum{}, and high off-band blocking density due to KDP (Fig.\,\ref{fig:Zband}).

\subsection{Alternative materials}
We are not aware any other materials than the KDP and its isomorphisms that absorb thermal IR photons as effectively while, being transparent in the NIR and bluer wavelengths. The heat absorbing glasses offered by several major glass manufacturers have high optical density only at wavelengths $>$2.7\,\mum{}, and significantly lower transmission in the NIR range. Compared to a typical heat absorbing glass, KDP offers a longer cutoff wavelength, and orders of magnitude higher optical density at wavelengths longer than 1.8\,\mum{}. Based on Table\,\ref{table:isomorphs}, we identify KDA and RDA as potential materials candidates for NIR shortpass filter, due to their lower than KDP Curie temperature, and suitable cutoff wavelength. We have not tested these materials, and they are not as easily available as KDP.

\subsection{Possible caveats}
KDP contains Potassium which has radioactive isotopes. The radioactive decay leads to particle events observable by the photo-detectors if not shielded by a cover glass. KDP is a birefringent material, and the cutoff wavelength $\lambda_{\rm SP}$ has a polarization dependence, which may affect some use cases. KDP is also water soluble \citep{CRCoptical}, and care should be taken handling it. Since we had no prior experience with the material, we kept the KDPs inside their vacuum bags until the measurement and installation in the instrument. We used standard nitrile gloves and face masks when handling the crystals with the expectation that the moisture from hands or breath could be sufficient to damage the uncoated optical surfaces. The instrument was closed and evacuated from air soon after mounting the crystals. We intend to do further cryogenic gluing and coating tests at a later date when the instrument is opened again and the crystals are going to be unmounted.

\section{Conclusion}\label{sec:conclusions}
We have shown that the KDP is an excellent thermal blocking filter and practical to use in an astronomical context, with limitation on its operating temperature, and birefringence. KDP offers good optical characteristics, and can strongly reduce atmospheric or instrumental thermal background.  We have demonstrated the use of a warm narrow band filter in conjunction with a 2.5\,\mum{} cutoff detector, which has been enabled by the high thermal IR absorbance of the KDP. While we cannot demonstrate a photometric gain in the warm narrowband over the J band, we measure the night sky surface brightness within the bandpass, and find it to be 18.5\,\magarcsec{}. KDP has the potential to become a mainstay in the SWIR astronomy as a heat absorbing, thermal emission blocking, shortpass filter material. It offers a way to improve the SWIR broad- and especially narrowband filter efficiency, and off-band blocking characteristics, can be used for enhancing performance of existing instrumentation suffering from thermal leaks, and has potential for realizing new instrument concepts.

\section*{Data availability}\label{sec:availability}
Data are available upon reasonable request from the corresponding author and the Nordic Optical Telescope's FITS-file archive. Software is available from the corresponding author upon reasonable request.
    
\begin{acknowledgments}
Cosmic Dawn Center (DAWN) is funded by the Danish National Research Foundation under grant DNRF140. We thank C. Pérez, T. Pursimo, G. Cox, P. Brandt, A. Kasikov, P. Galindo-Guil, M. Keniger, J. Saario, A. M. Kadela, and A. Henderson de la Fuente for their assistance at the NOT, Instituto de Astrofísica de Canarias (IAC) IACTEC optical laboratory for the Z and Y-band filter scans, and V. Pinter and A. E. T. Viitanen for their feedback on the manuscript.
\end{acknowledgments}

\begin{contribution}

MIA conceptualized the project. JPUF acquired hardware funding for the project. JKMV managed the project. JKMV, AAD, and SA recorded the observatory site data. JKMV and AAD processed and analyzed the data. ANS and MIA carried out the thermal analysis. DK and PS recorded the optical density data, and designed and manufactured the NB filter. JKMV wrote the original draft. All authors contributed equally to the review and editing.


\end{contribution}

%
\facilities{NOT (NOTCam)}


\software{Astropy \citep{astropy:2013,astropy:2018,astropy:2022},  
          Astroquery \citep{astroquery}, 
          Matplotlib \citep{matplotlib},
          Numpy \citep{numpy},
          Reproject\footnote{\url{https://github.com/astropy/reproject}}
          }


\newpage
\appendix
\restartappendixnumbering

\section{Instrument setup}
\mbox{}
\begin{figure*}[!ht]
    \centering
    \includegraphics[width=.95\textwidth]{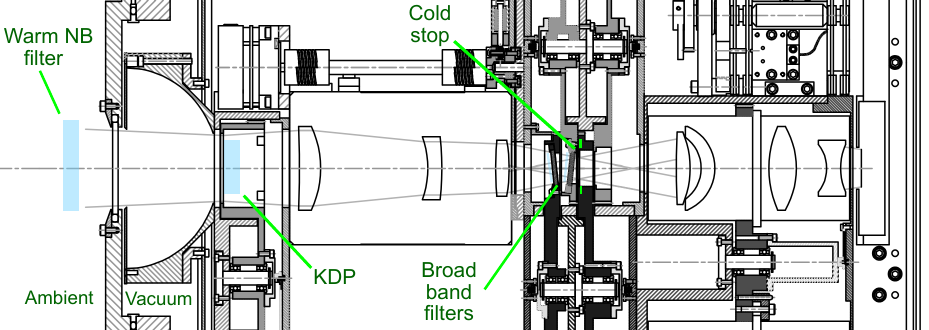}
    \caption{Schematic layout of the \mbox{NOTCam} optics, and the additional optical components used in this study. The warm NB filter was installed in the telescope converging beam, outside of the instruments' cryostat window. The telescope focus is placed right after \mbox{NOTCam}'s entrance baffle. The KDPs were installed in the aperture wheel placing them in the telescope focus. Broadband filters were installed in their standard location in the \mbox{NOTCam} parallel beam with a 5$\degr$ tilt respective to the optical axis.
    \label{fig:notcam}
    }
\end{figure*}   

\begin{figure*}[!ht]
    \gridline{\fig{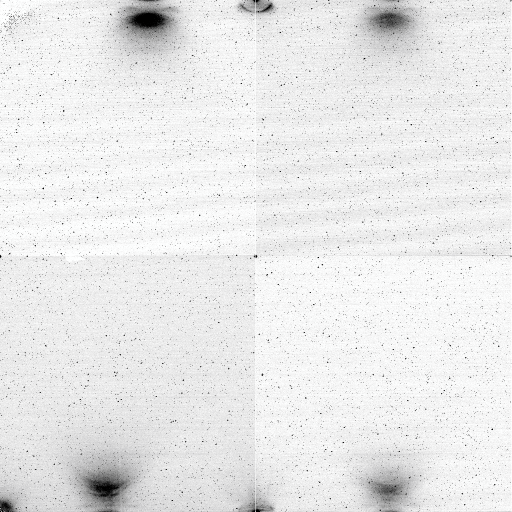}{0.3\textwidth}{(a)}
              \fig{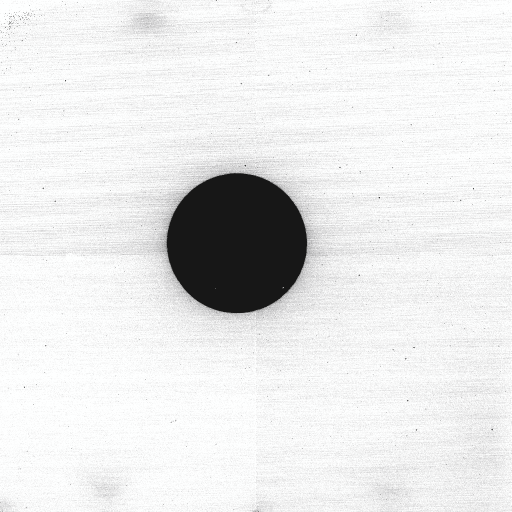}{0.3\textwidth}{(b)}
              \fig{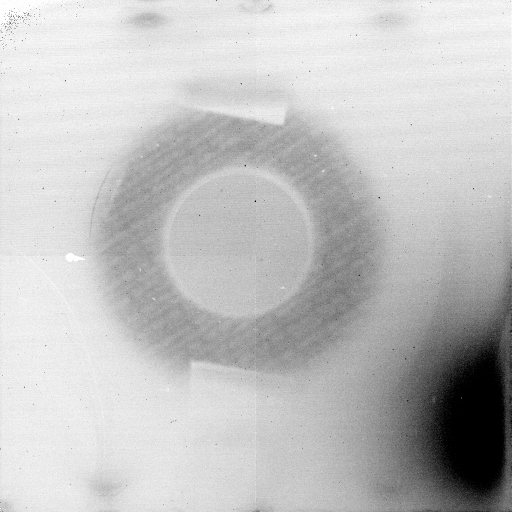}{0.3\textwidth}{(c)} 
             }
    \caption{Inverted grayscale illustration of detector illumination. (a) dark frame, (b) J-band sky image showing the restricted image circle when the fiberglass insulated KDP is placed  in the beam, and (c) K band integration showing a light leak from adjacent aperture, and partially transparent fiberglass holder. Integration times and color scales are not equal in the three panels.
    \label{fig:detector}
    }
\end{figure*}

\begin{figure*}[!ht]
    \gridline{\fig{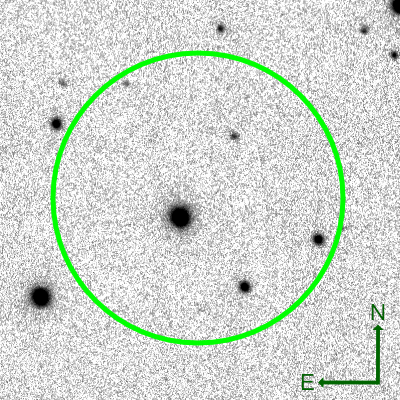}{0.37\textwidth}{(a)}
              \fig{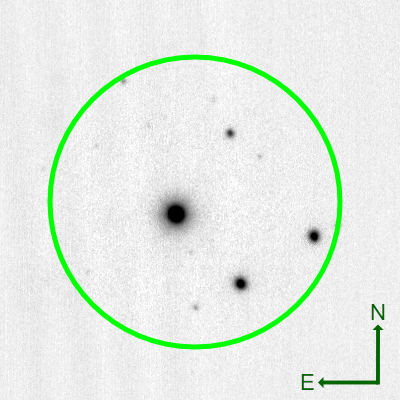}{0.37\textwidth}{(b)}
    }
    \caption{RL\,149 field observed without (a), and with (b) KDP in the \mbox{NOTCam} Z band. SNR increase is clear in the images with equal equal integration times, and equal logarithmic color scale. Size of KDP image circle is indicated by a green circle.
    \label{fig:zcomp} 
    }
\end{figure*}

\begin{figure*}[!ht]
    \gridline{\fig{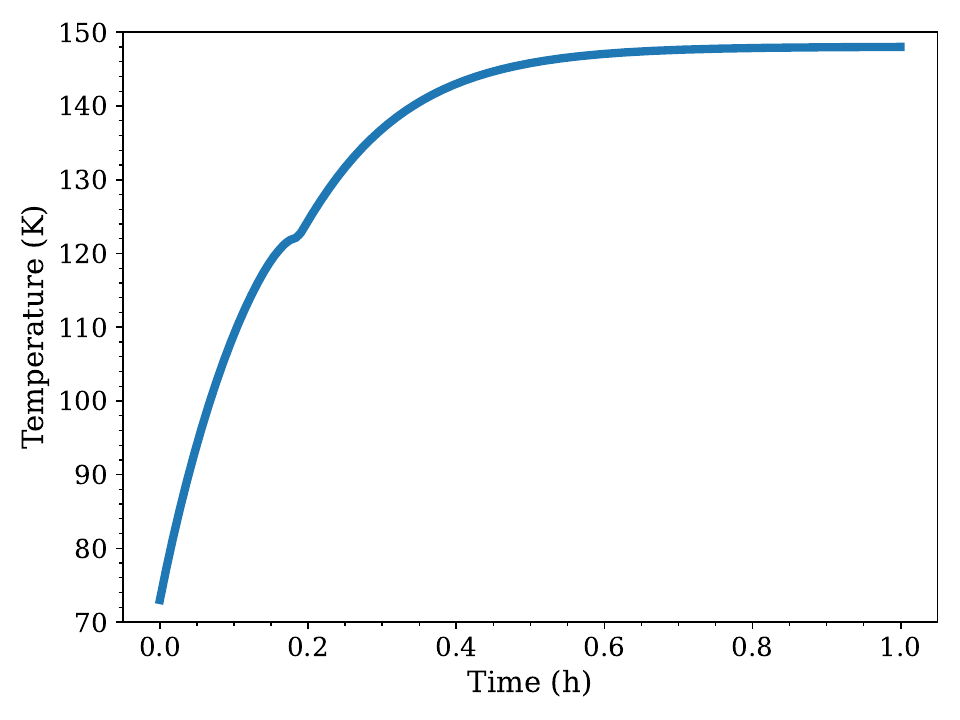}{0.4\textwidth}{(a)}
              \fig{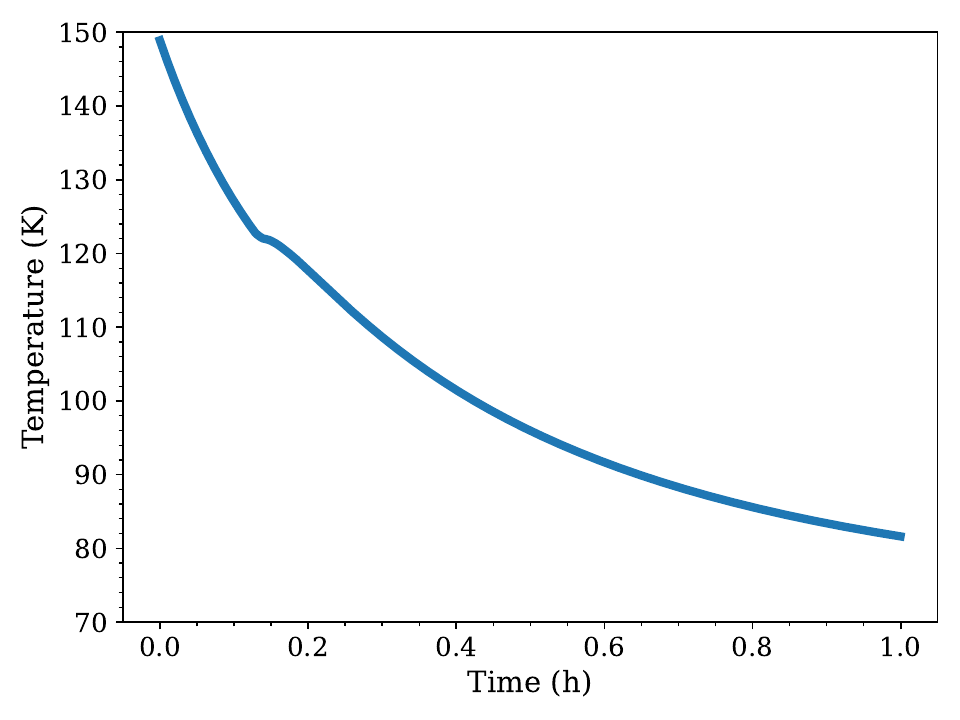}{0.4\textwidth}{(b)}
    }
    \caption{Calculated temperature of the 0.5\,mm KDP thick fiber glass insulated in the \mbox{NOTCam} aperture wheel, radiatively warming up when moved in the telescope beam (a), and cooling cool down when moving out of telescope beam (b). The heat capacity of KDP has an anomaly close to its $T_{\mathrm{C}}$ \citep{Stephenson44a}, causing a small kink in the temperature. 
    \label{fig:kdpinout} 
    }
\end{figure*}

\newpage
\section{KDP optical density data}
\begin{deluxetable*}{rlcrlcrlcrlcrlcrl}[!ht]
\tabletypesize{\scriptsize}
\caption{Optical density of a 1\,mm thick KDP measured with unpolarized light, \emph{continues} in Table\,\ref{table:od2}
\label{table:od1}
}

\tablehead{
    \colhead{$\lambda$ (\mum{})} & \colhead{OD} & \colhead{} & \colhead{$\lambda$ (\mum{})} & \colhead{OD} & \colhead{} & \colhead{$\lambda$ (\mum{})} & \colhead{OD} & \colhead{} & \colhead{$\lambda$ (\mum{})} & \colhead{OD} & \colhead{} & \colhead{$\lambda$ (\mum{})} & \colhead{OD} & \colhead{} & \colhead{$\lambda$ (\mum{})} & \colhead{OD}
}
\startdata
      &        &  &       &        &  &       &        &  &       &        &  &        &        &  &       &        \\
0.299 & 0.0028 &  & 0.467 & 0.0003 &  & 0.635 & 0.0001 &  & 0.803 & 0.0002 &  & 0.971  & 0.0008 &  & 1.139 & 0.0020 \\
0.302 & 0.0027 &  & 0.470 & 0.0001 &  & 0.638 & 0.0001 &  & 0.806 & 0.0003 &  & 0.974  & 0.0008 &  & 1.142 & 0.0021 \\
0.305 & 0.0025 &  & 0.473 & 0.0004 &  & 0.641 & 0.0003 &  & 0.809 & 0.0006 &  & 0.977  & 0.0007 &  & 1.145 & 0.0021 \\
0.308 & 0.0022 &  & 0.476 & 0.0004 &  & 0.644 & 0.0001 &  & 0.812 & 0.0005 &  & 0.980  & 0.0010 &  & 1.148 & 0.0019 \\
0.311 & 0.0021 &  & 0.479 & 0.0002 &  & 0.647 & 0.0001 &  & 0.815 & 0.0003 &  & 0.983  & 0.0009 &  & 1.151 & 0.0021 \\
0.314 & 0.0017 &  & 0.482 & 0.0001 &  & 0.650 & 0.0001 &  & 0.818 & 0.0002 &  & 0.986  & 0.0008 &  & 1.154 & 0.0022 \\
0.317 & 0.0017 &  & 0.485 & 0.0002 &  & 0.653 & 0.0003 &  & 0.821 & 0.0006 &  & 0.989  & 0.0007 &  & 1.157 & 0.0023 \\
0.320 & 0.0015 &  & 0.488 & 0.0001 &  & 0.656 & 0.0001 &  & 0.824 & 0.0007 &  & 0.992  & 0.0011 &  & 1.160 & 0.0022 \\
0.323 & 0.0011 &  & 0.491 & 0.0001 &  & 0.659 & 0.0001 &  & 0.827 & 0.0006 &  & 0.995  & 0.0009 &  & 1.163 & 0.0024 \\
0.326 & 0.0011 &  & 0.494 & 0.0002 &  & 0.662 & 0.0001 &  & 0.830 & 0.0002 &  & 0.998  & 0.0009 &  & 1.166 & 0.0024 \\
0.329 & 0.0009 &  & 0.497 & 0.0004 &  & 0.665 & 0.0003 &  & 0.833 & 0.0003 &  & 1.001  & 0.0009 &  & 1.169 & 0.0022 \\
0.332 & 0.0007 &  & 0.500 & 0.0005 &  & 0.668 & 0.0002 &  & 0.836 & 0.0003 &  & 1.004  & 0.0010 &  & 1.172 & 0.0022 \\
0.335 & 0.0007 &  & 0.503 & 0.0005 &  & 0.671 & 0.0002 &  & 0.839 & 0.0004 &  & 1.007  & 0.0010 &  & 1.175 & 0.0022 \\
0.338 & 0.0007 &  & 0.506 & 0.0007 &  & 0.674 & 0.0001 &  & 0.842 & 0.0002 &  & 1.010  & 0.0010 &  & 1.178 & 0.0025 \\
0.341 & 0.0004 &  & 0.509 & 0.0007 &  & 0.677 & 0.0002 &  & 0.845 & 0.0002 &  & 1.013  & 0.0011 &  & 1.181 & 0.0026 \\
0.344 & 0.0002 &  & 0.512 & 0.0004 &  & 0.680 & 0.0001 &  & 0.848 & 0.0004 &  & 1.016  & 0.0012 &  & 1.184 & 0.0026 \\
0.347 & 0.0001 &  & 0.515 & 0.0005 &  & 0.683 & 0.0001 &  & 0.851 & 0.0001 &  & 1.019  & 0.0013 &  & 1.187 & 0.0026 \\
0.350 & 0.0003 &  & 0.518 & 0.0001 &  & 0.686 & 0.0001 &  & 0.854 & 0.0001 &  & 1.022  & 0.0010 &  & 1.190 & 0.0027 \\
0.353 & 0.0004 &  & 0.521 & 0.0002 &  & 0.689 & 0.0001 &  & 0.857 & 0.0008 &  & 1.025  & 0.0011 &  & 1.193 & 0.0029 \\
0.356 & 0.0002 &  & 0.524 & 0.0005 &  & 0.692 & 0.0001 &  & 0.860 & 0.0006 &  & 1.028  & 0.0011 &  & 1.196 & 0.0030 \\
0.359 & 0.0001 &  & 0.527 & 0.0001 &  & 0.695 & 0.0001 &  & 0.863 & 0.0005 &  & 1.031  & 0.0010 &  & 1.199 & 0.0029 \\
0.362 & 0.0001 &  & 0.530 & 0.0001 &  & 0.698 & 0.0001 &  & 0.866 & 0.0005 &  & 1.034  & 0.0012 &  & 1.202 & 0.0031 \\
0.365 & 0.0002 &  & 0.533 & 0.0004 &  & 0.701 & 0.0002 &  & 0.869 & 0.0004 &  & 1.037  & 0.0011 &  & 1.205 & 0.0033 \\
0.368 & 0.0001 &  & 0.536 & 0.0004 &  & 0.704 & 0.0002 &  & 0.872 & 0.0006 &  & 1.040  & 0.0011 &  & 1.208 & 0.0035 \\
0.371 & 0.0001 &  & 0.539 & 0.0001 &  & 0.707 & 0.0001 &  & 0.875 & 0.0002 &  & 1.043  & 0.0011 &  & 1.211 & 0.0033 \\
0.374 & 0.0002 &  & 0.542 & 0.0001 &  & 0.710 & 0.0008 &  & 0.878 & 0.0003 &  & 1.046  & 0.0011 &  & 1.214 & 0.0035 \\
0.377 & 0.0001 &  & 0.545 & 0.0002 &  & 0.713 & 0.0001 &  & 0.881 & 0.0001 &  & 1.049  & 0.0010 &  & 1.217 & 0.0035 \\
0.380 & 0.0003 &  & 0.548 & 0.0003 &  & 0.716 & 0.0001 &  & 0.884 & 0.0004 &  & 1.052  & 0.0013 &  & 1.220 & 0.0036 \\
0.383 & 0.0004 &  & 0.551 & 0.0001 &  & 0.719 & 0.0001 &  & 0.887 & 0.0002 &  & 1.055  & 0.0012 &  & 1.223 & 0.0038 \\
0.386 & 0.0001 &  & 0.554 & 0.0001 &  & 0.722 & 0.0002 &  & 0.890 & 0.0003 &  & 1.058  & 0.0014 &  & 1.226 & 0.0040 \\
0.389 & 0.0001 &  & 0.557 & 0.0001 &  & 0.725 & 0.0001 &  & 0.893 & 0.0002 &  & 1.061  & 0.0011 &  & 1.229 & 0.0041 \\
0.392 & 0.0002 &  & 0.560 & 0.0001 &  & 0.728 & 0.0001 &  & 0.896 & 0.0006 &  & 1.064  & 0.0013 &  & 1.232 & 0.0042 \\
0.395 & 0.0001 &  & 0.563 & 0.0001 &  & 0.731 & 0.0001 &  & 0.899 & 0.0004 &  & 1.067  & 0.0014 &  & 1.235 & 0.0045 \\
0.398 & 0.0002 &  & 0.566 & 0.0004 &  & 0.734 & 0.0001 &  & 0.902 & 0.0004 &  & 1.070  & 0.0012 &  & 1.238 & 0.0046 \\
0.401 & 0.0001 &  & 0.569 & 0.0002 &  & 0.737 & 0.0002 &  & 0.905 & 0.0003 &  & 1.073  & 0.0014 &  & 1.241 & 0.0047 \\
0.404 & 0.0002 &  & 0.572 & 0.0001 &  & 0.740 & 0.0001 &  & 0.908 & 0.0006 &  & 1.076  & 0.0014 &  & 1.244 & 0.0048 \\
0.407 & 0.0001 &  & 0.575 & 0.0001 &  & 0.743 & 0.0001 &  & 0.911 & 0.0004 &  & 1.079  & 0.0014 &  & 1.247 & 0.0050 \\
0.410 & 0.0001 &  & 0.578 & 0.0001 &  & 0.746 & 0.0001 &  & 0.914 & 0.0004 &  & 1.082  & 0.0015 &  & 1.250 & 0.0051 \\
0.413 & 0.0001 &  & 0.581 & 0.0004 &  & 0.749 & 0.0004 &  & 0.917 & 0.0005 &  & 1.085  & 0.0014 &  & 1.253 & 0.0054 \\
0.416 & 0.0002 &  & 0.584 & 0.0003 &  & 0.752 & 0.0001 &  & 0.920 & 0.0007 &  & 1.088  & 0.0013 &  & 1.256 & 0.0057 \\
0.419 & 0.0001 &  & 0.587 & 0.0001 &  & 0.755 & 0.0001 &  & 0.923 & 0.0006 &  & 1.091  & 0.0016 &  & 1.259 & 0.0057 \\
0.422 & 0.0002 &  & 0.590 & 0.0001 &  & 0.758 & 0.0001 &  & 0.926 & 0.0007 &  & 1.094  & 0.0015 &  & 1.262 & 0.0058 \\
0.425 & 0.0001 &  & 0.593 & 0.0001 &  & 0.761 & 0.0001 &  & 0.929 & 0.0005 &  & 1.097  & 0.0017 &  & 1.265 & 0.0059 \\
0.428 & 0.0005 &  & 0.596 & 0.0001 &  & 0.764 & 0.0001 &  & 0.932 & 0.0009 &  & 1.100  & 0.0017 &  & 1.268 & 0.0063 \\
0.431 & 0.0003 &  & 0.599 & 0.0001 &  & 0.767 & 0.0001 &  & 0.935 & 0.0004 &  & 1.103  & 0.0017 &  & 1.271 & 0.0066 \\
0.434 & 0.0001 &  & 0.602 & 0.0002 &  & 0.770 & 0.0002 &  & 0.938 & 0.0006 &  & 1.106  & 0.0016 &  & 1.274 & 0.0069 \\
0.437 & 0.0002 &  & 0.605 & 0.0001 &  & 0.773 & 0.0001 &  & 0.941 & 0.0005 &  & 1.109  & 0.0017 &  & 1.277 & 0.0072 \\
0.440 & 0.0002 &  & 0.608 & 0.0001 &  & 0.776 & 0.0001 &  & 0.944 & 0.0008 &  & 1.112  & 0.0018 &  & 1.280 & 0.0074 \\
0.443 & 0.0001 &  & 0.611 & 0.0001 &  & 0.779 & 0.0002 &  & 0.947 & 0.0007 &  & 1.115  & 0.0017 &  & 1.283 & 0.0077 \\
0.446 & 0.0001 &  & 0.614 & 0.0004 &  & 0.782 & 0.0002 &  & 0.950 & 0.0008 &  & 1.118  & 0.0019 &  & 1.286 & 0.0079 \\
0.449 & 0.0003 &  & 0.617 & 0.0004 &  & 0.785 & 0.0003 &  & 0.953 & 0.0006 &  & 1.121  & 0.0019 &  & 1.289 & 0.0082 \\
0.452 & 0.0006 &  & 0.620 & 0.0002 &  & 0.788 & 0.0001 &  & 0.956 & 0.0008 &  & 1.124  & 0.0019 &  & 1.292 & 0.0084 \\
0.455 & 0.0003 &  & 0.623 & 0.0002 &  & 0.791 & 0.0001 &  & 0.959 & 0.0006 &  & 1.127  & 0.0018 &  & 1.295 & 0.0089 \\
0.458 & 0.0002 &  & 0.626 & 0.0001 &  & 0.794 & 0.0004 &  & 0.962 & 0.0008 &  & 1.130  & 0.0018 &  & 1.298 & 0.0093 \\
0.461 & 0.0003 &  & 0.629 & 0.0001 &  & 0.797 & 0.0006 &  & 0.965 & 0.0008 &  & 1.133  & 0.0019 &  & 1.301 & 0.0095 \\
0.464 & 0.0006 &  & 0.632 & 0.0001 &  & 0.800 & 0.0004 &  & 0.968 & 0.0009 &  & 1.136  & 0.0019 &  & 1.304 & 0.0100 \\
      &        &  &       &        &  &       &        &  &       &        &  &        &        &  &       &        \\
\enddata
\tablecomments{Fresnel losses corrected. Scans of 4\,mm ($<$1.86\,\mum{}) and 0.5\,mm ($>$1.86\,\mum{}) thick samples have been combined into a 1\,mm equivalent optical density.
}
\end{deluxetable*}

\begin{deluxetable*}{rlcrlcrlcrlcrlcrl}[!ht]
\tabletypesize{\scriptsize}
\caption{\emph{continued} Optical density of a 1\,mm thick KDP measured with unpolarized light
\label{table:od2}
}

\tablehead{
    \colhead{$\lambda$ (\mum{})} & \colhead{OD} & \colhead{} & \colhead{$\lambda$ (\mum{})} & \colhead{OD} & \colhead{} & \colhead{$\lambda$ (\mum{})} & \colhead{OD} & \colhead{} & \colhead{$\lambda$ (\mum{})} & \colhead{OD} & \colhead{} & \colhead{$\lambda$ (\mum{})} & \colhead{OD} & \colhead{} & \colhead{$\lambda$ (\mum{})} & \colhead{OD}
}
\startdata
      &        &  &       &        &  &       &        &  &       &        &  &       &        &  &       &        \\
1.307 & 0.0104 &  & 1.478 & 0.0649 &  & 1.646 & 0.2952 &  & 1.814 & 1.1399 &  & 1.982 & 3.2434 &  & 2.150 & 5.2101 \\
1.310 & 0.0108 &  & 1.481 & 0.0669 &  & 1.649 & 0.3028 &  & 1.817 & 1.1659 &  & 1.985 & 3.2774 &  & 2.153 & 5.2727 \\
1.313 & 0.0110 &  & 1.484 & 0.0687 &  & 1.652 & 0.3107 &  & 1.820 & 1.1889 &  & 1.988 & 3.3078 &  & 2.156 & 5.2783 \\
1.316 & 0.0116 &  & 1.487 & 0.0707 &  & 1.655 & 0.3188 &  & 1.823 & 1.2023 &  & 1.991 & 3.3062 &  & 2.159 & 5.3583 \\
1.319 & 0.0121 &  & 1.490 & 0.0727 &  & 1.658 & 0.3273 &  & 1.826 & 1.2082 &  & 1.994 & 3.3412 &  & 2.162 & 5.3959 \\
1.322 & 0.0125 &  & 1.493 & 0.0748 &  & 1.661 & 0.3356 &  & 1.829 & 1.2528 &  & 1.997 & 3.3750 &  & 2.165 & 5.4345 \\
1.325 & 0.0130 &  & 1.496 & 0.0769 &  & 1.664 & 0.3444 &  & 1.832 & 1.2790 &  & 2.000 & 3.4118 &  & 2.168 & 5.5310 \\
1.328 & 0.0133 &  & 1.499 & 0.0790 &  & 1.667 & 0.3637 &  & 1.835 & 1.2924 &  & 2.003 & 3.4846 &  & 2.171 & 5.5510 \\
1.331 & 0.0137 &  & 1.502 & 0.0813 &  & 1.670 & 0.3657 &  & 1.838 & 1.3171 &  & 2.006 & 3.5629 &  & 2.174 & 5.5732 \\
1.334 & 0.0142 &  & 1.505 & 0.0838 &  & 1.673 & 0.3717 &  & 1.841 & 1.3574 &  & 2.009 & 3.5559 &  & 2.177 & 5.6370 \\
1.337 & 0.0149 &  & 1.508 & 0.0861 &  & 1.676 & 0.3809 &  & 1.844 & 1.3359 &  & 2.012 & 3.6307 &  & 2.180 & 5.6492 \\
1.340 & 0.0155 &  & 1.511 & 0.0887 &  & 1.679 & 0.3910 &  & 1.847 & 1.4140 &  & 2.015 & 3.6527 &  & 2.183 & 5.7146 \\
1.346 & 0.0167 &  & 1.514 & 0.0913 &  & 1.682 & 0.4009 &  & 1.850 & 1.4653 &  & 2.018 & 3.7075 &  & 2.186 & 5.8368 \\
1.349 & 0.0172 &  & 1.517 & 0.0938 &  & 1.685 & 0.4117 &  & 1.853 & 1.4399 &  & 2.021 & 3.7453 &  & 2.189 & 5.8118 \\
1.352 & 0.0177 &  & 1.520 & 0.0965 &  & 1.688 & 0.4221 &  & 1.856 & 1.6654 &  & 2.024 & 3.7523 &  & 2.192 & 5.8286 \\
1.355 & 0.0185 &  & 1.523 & 0.0993 &  & 1.691 & 0.4331 &  & 1.859 & 1.6396 &  & 2.027 & 3.8201 &  & 2.195 & 6.0248 \\
1.358 & 0.0191 &  & 1.526 & 0.1024 &  & 1.694 & 0.4442 &  & 1.862 & 1.5946 &  & 2.030 & 3.8257 &  & 2.198 & 6.0985 \\
1.361 & 0.0195 &  & 1.529 & 0.1053 &  & 1.697 & 0.4557 &  & 1.865 & 1.5466 &  & 2.033 & 3.8843 &  & 2.201 & 6.0855 \\
1.364 & 0.0201 &  & 1.532 & 0.1083 &  & 1.700 & 0.4674 &  & 1.868 & 1.6601 &  & 2.036 & 3.8609 &  & 2.204 & 6.0411 \\
1.367 & 0.0210 &  & 1.535 & 0.1113 &  & 1.703 & 0.4795 &  & 1.871 & 1.8927 &  & 2.039 & 3.9120 &  & 2.207 & 6.1293 \\
1.370 & 0.0218 &  & 1.538 & 0.1147 &  & 1.706 & 0.4922 &  & 1.874 & 1.9059 &  & 2.042 & 4.0072 &  & 2.210 & 6.2227 \\
1.373 & 0.0223 &  & 1.541 & 0.1179 &  & 1.709 & 0.5046 &  & 1.877 & 1.8317 &  & 2.045 & 4.0850 &  & 2.213 & 6.3057 \\
1.376 & 0.0230 &  & 1.544 & 0.1212 &  & 1.712 & 0.5174 &  & 1.880 & 1.8347 &  & 2.048 & 4.0586 &  & 2.216 & 6.3359 \\
1.379 & 0.0239 &  & 1.547 & 0.1247 &  & 1.715 & 0.5305 &  & 1.883 & 1.8149 &  & 2.051 & 4.1152 &  & 2.219 & 6.4513 \\
1.382 & 0.0247 &  & 1.550 & 0.1282 &  & 1.718 & 0.5438 &  & 1.886 & 1.8455 &  & 2.054 & 4.1394 &  & 2.222 & 6.5345 \\
1.385 & 0.0255 &  & 1.553 & 0.1319 &  & 1.721 & 0.5579 &  & 1.889 & 1.9231 &  & 2.057 & 4.0972 &  & 2.225 & 6.4211 \\
1.388 & 0.0263 &  & 1.556 & 0.1357 &  & 1.724 & 0.5716 &  & 1.892 & 1.8967 &  & 2.060 & 4.1658 &  & 2.228 & 6.4486 \\
1.391 & 0.0271 &  & 1.559 & 0.1396 &  & 1.727 & 0.5858 &  & 1.895 & 1.8359 &  & 2.063 & 4.2442 &  & 2.231 & 6.6304 \\
1.394 & 0.0282 &  & 1.562 & 0.1433 &  & 1.730 & 0.6002 &  & 1.898 & 1.9807 &  & 2.066 & 4.2802 &  & 2.234 & 6.7272 \\
1.397 & 0.0293 &  & 1.565 & 0.1473 &  & 1.733 & 0.6150 &  & 1.901 & 2.0205 &  & 2.069 & 4.3018 &  & 2.237 & 6.7992 \\
1.400 & 0.0301 &  & 1.568 & 0.1512 &  & 1.736 & 0.6302 &  & 1.904 & 2.0490 &  & 2.072 & 4.3405 &  & 2.240 & 6.8138 \\
1.403 & 0.0309 &  & 1.571 & 0.1554 &  & 1.739 & 0.6455 &  & 1.907 & 2.1578 &  & 2.075 & 4.3691 &  & 2.243 & 6.8610 \\
1.406 & 0.0319 &  & 1.574 & 0.1596 &  & 1.742 & 0.6613 &  & 1.910 & 2.1142 &  & 2.078 & 4.3937 &  & 2.246 & 6.8878 \\
1.409 & 0.0329 &  & 1.577 & 0.1639 &  & 1.745 & 0.6776 &  & 1.913 & 2.1506 &  & 2.081 & 4.3751 &  & 2.249 & 6.9478 \\
1.412 & 0.0339 &  & 1.580 & 0.1685 &  & 1.748 & 0.6940 &  & 1.916 & 2.0992 &  & 2.084 & 4.3999 &  & 2.252 & 7.0958 \\
1.415 & 0.0350 &  & 1.583 & 0.1730 &  & 1.751 & 0.7108 &  & 1.919 & 2.2050 &  & 2.087 & 4.4605 &  & 2.255 & 7.0162 \\
1.418 & 0.0363 &  & 1.586 & 0.1777 &  & 1.754 & 0.7280 &  & 1.922 & 2.3220 &  & 2.090 & 4.4735 &  & 2.258 & 7.0443 \\
1.421 & 0.0373 &  & 1.589 & 0.1825 &  & 1.757 & 0.7453 &  & 1.925 & 2.4336 &  & 2.093 & 4.4799 &  & 2.261 & 7.2821 \\
1.424 & 0.0384 &  & 1.592 & 0.1872 &  & 1.760 & 0.7632 &  & 1.928 & 2.4488 &  & 2.096 & 4.5179 &  & 2.264 & 7.2323 \\
1.427 & 0.0397 &  & 1.595 & 0.1922 &  & 1.763 & 0.7811 &  & 1.931 & 2.4882 &  & 2.099 & 4.5217 &  & 2.267 & 7.4013 \\
1.430 & 0.0408 &  & 1.598 & 0.1970 &  & 1.766 & 0.7995 &  & 1.934 & 2.5374 &  & 2.102 & 4.6067 &  & 2.270 & 7.4909 \\
1.433 & 0.0420 &  & 1.601 & 0.2024 &  & 1.769 & 0.8182 &  & 1.937 & 2.6211 &  & 2.105 & 4.6508 &  & 2.273 & 7.5017 \\
1.436 & 0.0433 &  & 1.604 & 0.2074 &  & 1.772 & 0.8378 &  & 1.940 & 2.6261 &  & 2.108 & 4.6394 &  & 2.276 & 7.2635 \\
1.439 & 0.0446 &  & 1.607 & 0.2127 &  & 1.775 & 0.8574 &  & 1.943 & 2.7201 &  & 2.111 & 4.7220 &  & 2.279 & 7.2751 \\
1.442 & 0.0461 &  & 1.610 & 0.2180 &  & 1.778 & 0.8777 &  & 1.946 & 2.7295 &  & 2.114 & 4.7638 &  & 2.282 & 7.2751 \\
1.445 & 0.0474 &  & 1.613 & 0.2237 &  & 1.781 & 0.8981 &  & 1.949 & 2.7841 &  & 2.117 & 4.8012 &  & 2.285 & 7.4599 \\
1.448 & 0.0488 &  & 1.616 & 0.2292 &  & 1.784 & 0.9183 &  & 1.952 & 2.7859 &  & 2.120 & 4.8308 &  & 2.288 & 7.2956 \\
1.451 & 0.0504 &  & 1.619 & 0.2349 &  & 1.787 & 0.9379 &  & 1.955 & 2.8031 &  & 2.123 & 4.7988 &  & 2.291 & 7.4444 \\
1.454 & 0.0517 &  & 1.622 & 0.2410 &  & 1.790 & 0.9578 &  & 1.958 & 2.9011 &  & 2.126 & 4.8946 &  & 2.294 & 7.6386 \\
1.457 & 0.0530 &  & 1.625 & 0.2472 &  & 1.793 & 0.9776 &  & 1.961 & 2.9535 &  & 2.129 & 4.9920 &  & 2.297 & 7.5730 \\
1.460 & 0.0545 &  & 1.628 & 0.2536 &  & 1.796 & 0.9969 &  & 1.964 & 3.0751 &  & 2.132 & 4.9930 &  & 2.300 & 7.9230 \\
1.463 & 0.0563 &  & 1.631 & 0.26 0 &  & 1.799 & 1.0161 &  & 1.967 & 2.9675 &  & 2.135 & 5.0087 &  & 2.303 & 7.9608 \\
1.466 & 0.0580 &  & 1.634 & 0.2667 &  & 1.802 & 1.0636 &  & 1.970 & 3.0119 &  & 2.138 & 5.0393 &  & 2.306 & 7.8872 \\
1.469 & 0.0596 &  & 1.637 & 0.2734 &  & 1.805 & 1.0805 &  & 1.973 & 3.0760 &  & 2.141 & 5.0473 &  & 2.309 & 7.9740 \\
1.472 & 0.0613 &  & 1.640 & 0.2805 &  & 1.808 & 1.1011 &  & 1.976 & 3.1396 &  & 2.144 & 5.0733 &  & 2.312 & 8.0710 \\
1.475 & 0.0630 &  & 1.643 & 0.2879 &  & 1.811 & 1.1207 &  & 1.979 & 3.1690 &  & 2.147 & 5.1257 &  & 2.315 & 8.1818 \\
      &        &  &       &        &  &       &        &  &       &        &  &       &        &  &       &        \\
\enddata
\tablecomments{Fresnel losses corrected. Scans of 4\,mm ($<$1.86\,\mum{}) and 0.5\,mm ($>$1.86\,\mum{}) thick samples have been combined into a 1\,mm equivalent optical density.
}
\end{deluxetable*}


\newpage
\bibliography{kdp_phot}{}
\bibliographystyle{aasjournalv7}



\end{document}